%% file: paper.tex
\pdfoutput=1

\documentclass[11pt]{article}

\input{front_matter}

\input{incl}

\begin{document}
\date{}

\title{Calibrating hypersonic turbulence flow models with the HIFiRE-1
  experiment using data-driven machine-learned models}

\author{Kenny Chowdhary\thanks{Extreme-scale Data Science and Analytics Department, Sandia National Laboratories, Livermore, CA 94550 (kchowdh@sandia.gov, ckhoang@sandia.gov, jairay@sandia.gov). Sandia National Laboratories is a multimission laboratory managed and operated by National Technology and Engineering Solutions of Sandia, LLC, a wholly owned subsidiary of Honeywell International, Inc., for the U.S. Department of Energy's National Nuclear Security Administration under contract DE-NA-0003525. }
\and 
Chi Hoang\footnotemark[1]
\and 
Kookjin Lee\footnotemark \thanks{School of Computing, Informatics, and Decision Systems Engineering, Arizona State University, Tempe, AZ 85281 (kookjin.lee@asu.edu)}
\and 
Jaideep Ray\footnotemark[1] }

\maketitle


\begin{abstract}
  In this paper we study the efficacy of combining machine-learning methods with
  projection-based model reduction techniques for creating data-driven surrogate models of
  computationally expensive, high-fidelity physics models. Such surrogate models are
  essential for many-query applications e.g., engineering design optimization and
  parameter estimation, where it is necessary to invoke the high-fidelity model
  sequentially, many times. Surrogate models are usually constructed for individual scalar
  quantities. However there are scenarios where a spatially varying field needs to be
  modeled as a function of the model's input parameters. We develop a method to do so,
  using projections to represent spatial variability while a machine-learned model
  captures the dependence of the model's response on the inputs. The method is
  demonstrated on modeling the heat flux and pressure on the surface of the HIFiRE-1
  geometry in a Mach 7.16 turbulent flow. The surrogate model is then used to perform
  Bayesian estimation of freestream conditions and parameters of the SST (Shear Stress
  Transport) turbulence model embedded in the high-fidelity (Reynolds-Averaged
  Navier-Stokes) flow simulator, using wind-tunnel data. The paper provides the first-ever
  Bayesian calibration of a turbulence model for complex hypersonic turbulent flows. We
  find that the primary issues in estimating the SST model parameters are the limited
  information content of the heat flux and pressure measurements and the large model-form
  error encountered in a certain part of the flow.
\end{abstract}



\section{Introduction}


In this paper, we develop and test a method to create surrogate models that can
approximate spatially varying responses (i.e., fields) generated by a high-fidelity
computational model (usually a system of partial differential equations) e.g., an
engineering simulator. The method will be demonstrated in the context of hypersonic
turbulent flow solutions over a realistic engineering geometry with shocks, boundary
layers, flow separation, and reattachment. Surrogate models are essential for
many-query applications e.g., design optimization or inverse problems where the
computational model has to be invoked repeatedly and sequentially, and for the prediction
of quantities of interest (henceforth \emph{QoI}), as a function of the computational
model's inputs. In this paper, we will demonstrate the surrogate models to calibrate a
turbulence model using data from a hypersonic wind-tunnel experiment, a process that will
require us to simulate a turbulent flow many times. Models for predicting a single scalar
variable have long existed (see Ref.~\cite{20aa3a,18ya3a,17sm3a} for reviews), as well as
for scalar-valued fields (see Ref.~\cite{15ac4a} for a review). A review of surrogate
modeling for aerodynamic applications can be found in Ref.~\cite{21kz2a}.

Swischuk et al.~\cite{19sm4a} describe an alternative way of surrogate modeling fields,
where they use some basic knowledge of the behavior of the fields in question to
significantly simplify the architecture (and therefore the training) of the surrogate
model. They realized that the spatial correlation in the fields persist and do not vary
erratically as the model inputs change, and thus the modeling could admit a
``separation-of-variables'' approach. They modeled the spatial variation of the field
using a basis set obtained by the proper orthogonal decomposition (POD) of a training
dataset of fields, and captured the dependence of the weights/coefficients of the bases on
the model inputs via machine-learning (ML) techniques. They found that simple ML methods
such as polynomial regressions were equal to, or better than, complex methods (such as
neural networks), which simplified the training of the models, and reduced requirements on
the size of the training dataset (TD). Hoang et al.~\cite{hoang2021projection} performed
the same study for unsteady (time-dependent) problems and arrived at much the same
conclusions. In both these foundational studies, the problems considered were
``idealized'' - the fields were smooth, their dependence on model parameters relatively
benign (though nonlinear) and the number of model input parameters less than half-a-dozen.

In this paper, we investigate whether a difficult and realistic engineering problem can be
addressed using the ``separation-of-variables'' approach to surrogate modeling, and
whether the usefulness of simple ML techniques still holds. We seek to construct surrogate
models for the heat-flux and pressure fields on the HIFiRE-1  geometry when placed in a
$\Minf = 7.16$ hypersonic flow in the LENS-I wind-tunnel (see Ref.~\cite{08wm4a,08mw4a} for a
description of the experiment and modeling effort). In our study, the high-fidelity
engineering simulator is a Reynolds-Averaged Navier--Stokes (RANS) model~\cite{06wd1a}
with Menter's SST turbulence model~\cite{94mf1a} embedded in it. Each simulation takes
about 384 CPU-hours\footnote{Each simulation is run using 128 2.3 GHz Intel Xeon Gold
  processors (4 nodes, each with 32 cores) for approximately 3 hours using Sandia's high
  performance computing resources.} to converge to steady state. The QoIs (heat-flux and
pressure) are obtained on the surface grid of the HIFiRE-1  geometry with 2170 grid
points. The RANS model has 12 uncertain parameters - freestream density, temperature and
velocity, as well as 9 SST turbulence model parameters - which form the input vector of
the surrogate model. The hypersonic flow contains discontinuities (shock waves), regions
of intense gradients (turbulent boundary layers) and a flow separation zone on the
HIFiRE-1  geometry. It is expected to pose a realistic challenge for surrogate modeling.

The usefulness of the surrogate model will be demonstrated by calibrating the 12 uncertain
parameters to wind-tunnel data (heat-flux and pressure measurements on the HIFiRE-1 
surface) The calibration will be Bayesian i.e., we will develop a 12-dimensional joint
probability density function (JPDF) over the uncertain parameters to capture the calibrated
values, as well as the uncertainty in them, due to a finite number of noisy measurements
and the shotcomings of the RANS model (i.e., model-form error).

This paper has two main contributions.  Firstly, we provide the first comprehensive and
comparative study of data-driven ROM-based surrogate model construction for a complex
realistic engineering application in hypersonic flows. In our case, the bulk of the
complexity is expected to arise from the ML models that represent the influence of the
freestream values and the nonlinearities engendered by the SST turbulence model. Surrogate
models that leverage knowledge of the physical phenomena to simplify their architecture,
training and TD requirements have their obvious attractions, especially when
computationally expensive models have to be run thousands of time to populate a
high-dimensional parameter space.

Our second contribution is the SST model calibrated to measurements from a 2D hypersonic
flow experiment. RANS models are approximate and often need to be calibrated to
experimental measurements from flows similar to their final use-case (e.g., in hypersonic
flows) to be predictive. To date, turbulence models calibrated to wind-tunnel data have
been limited to low-speed flows (incompressible~\cite{14gg3a} and
transonic~\cite{18rd5a}), and it is unclear whether the approximations inherent in RANS
will even allow the estimation of SST parameters with any degree of accuracy, given
separated hypersonic flows over the HIFiRE-1  geometry. While there have been attempts to
calibrate turbulence models in hypersonic flows~\cite{19zf2a}, they are limited to 1D
(flat-plate boundary layer) problems where many turbulent processes are absent. In
contrast, we provide a methodology, heavily reliant on surrogate modeling, that can be
used to calibrate models with data obtained from experiments that closely resemble actual
flight conditions.  We also provide the model that results from it, complete with
error/uncertainty estimates. Such a turbulence model, customized to hypersonic flows,
does not exist in aerospace engineering literature.

The paper is organized as follows.  In Sec.~\ref{sec:litrev}, we review existing
literature on surrogate modeling of fields and the state-of-the-art in turbulence model
calibration.  In Sec.~\ref{sec:prob} we describe the LENS-I experimental data, the SPARC
high-fidelity flow simulator, and the setup for model calibration. In
Sec.~\ref{sec:surrogate} we describe the construction of the surrogate model using
dimension reduction and different types of machine learning regressors, whose performance
is then evaluated in Sec.~\ref{sec:rom}. Sec.~\ref{sec:calib} contains the formulation and
results of the Bayesian calibration problem. Summary and conclusions are in
Sec.~\ref{sec:concl}.

\section{Literature review}
\label{sec:litrev}

\subsection{Surrogate models for fields}
\label{sec:surrmod}
Surrogate modeling of high-fidelity models is a mature topic and contemporary reviews of
the field can be found in Refs.~\cite{15ac4a,20aa3a,21vs2a}. We limit ourselves to
surrogate models that output spatial or spatiotemporal fields (see a review in
Ref.~\cite{15ac4a}). The process involves generating a large number of instances of the
field by executing the high-fidelity model repeatedly for different inputs (a vector of
independent scalar variables) and archiving the input-output pairs as TD. The TD is then
used to compute an orthogonal basis set, usually via POD~\cite{08my2a}, though Krylov
subspace bases~\cite{89dw2a,90wd3a} and Fourier bases~\cite{05wm2a,08gw2a} too have been
explored. The output field is then represented using a weighted linear combination of the
bases, with the (short) vector of weights serving as a low-dimensional representation of
the field. Weights, individually or as a vector, are then modeled as a function of the
uncertain inputs (of the high-fidelity model) via conventional data-driven methods. The
oldest example of such an approach seems to be Ref.~\cite{ly2001modeling}, where
Rayleigh-B\'{e}nard convection was modeled in this fashion, using cubic spline
interpolators to model the bases' coefficients (or weights). Gaussian Process models also
have been extensively used~\cite{higdon2008computer, audouze2009reduced,
  audouze2013nonintrusive, wirtz2015surrogate}, and there have been investigations into
using self-organizing maps coupled with local response
surfaces~\cite{mainini2015surrogate}. Neural networks can serve as universal approximators
and consequently Ref.~\cite{ulu2016data, hesthaven2018non} explore their use as a mapping
between inputs and the coefficients. In transient problems with a spatial component, POD
is often used to reduce spatial dimensionality while the time-evolving coefficients are
modeled as a dynamical system, using neural ordinary differential equations (NODE) and
recurrent neural networks~\cite{20mm6a,19rp5a,18wx6a}. Dimensionality reduction of a
spatial field $X$ can also be accomplished in a nonlinear manner using a neural net
encoder-decoder framework and used to predict a different, dependent spatial field $Y$ via
image-to-image regression~\cite{19mz5a}. In case of a spatiotemporal field, the
time-evolution of the field in the dimensionality-reduced encoded (or latent) space has
been modeled using a parameterized NODE~\cite{21lp2a}.  It is also possible to generate
the TD in an ``intelligent'' manner, sampling the input space where information on the
input-output relationship is desired~\cite{chen2018greedy}.  There are studies which
explore the benefits of various surrogate modeling techniques~\cite{17fs5a,17dm3a,19ll4a}
for a given problem, including in fluid mechanics~\cite{swischuk2019projection}.

Surrogate models have also been used in compressible aerothermodynamics (hypersonic
flows), often to approximate temperatures, pressures and heating on wings and control
surfaces~\cite{19cn5a,21dg4a}. They have been used within the context of
aerothermoelasticity studies~\cite{15cl4a,17cz4a,19cz2a,12cm2a,19xj4a} or to design
thermal protective systems~\cite{21zy5a}. In these surrogate models, the \emph{spatial}
variation of the fields are captured using POD bases, though a few studies have used
kriging~\cite{15cl4a,21dg4a,19xj4a}. In case of POD, the coefficient of the POD bases are
modeled as a function of the environment (e.g., Mach number, altitude etc.) using kriging,
radial basis functions or Chebyshev polynomials. The training dataset is generated using
computational fluid dynamics (CFD) simulators to span over a parameter space ranging from
two to eigth dimensions, consisting of Mach number (or speed), altitude and a host of
parameters describing the attitude of the vehicle and its control surfaces e.g., angle of
attack, roll angle etc. Comparisons between various surrogate modeling techniques for
hypersonic flow fields can be found in Refs.~\cite{19cn5a,15cl4a,12cm2a,19xj4a}. In these
studies the dataset for training the surrogate model generally contained
${\rm O(10) - O(10^2)}$ examples. In contrast, Ref.~\cite{21zy5a} models the
\emph{temporal} variation of temperature under the thermal protection system as a
hypersonic vehicle executes an uncertain trajectory using Karhunen-Lo\`{e}ve bases. Apart
from the three parameters that governed the uncertain trajectory, the study also
considered 18 uncertain parameters describing the material properties of the thermal
protective system. The training dataset had about 400 examples.
  
\subsection{Calibration of turbulence models}
\label{sec:calibturbmod}

Data-driven turbulence modeling has mostly targeted closures in RANS equations, though
some work has been done for Large Eddy Simulations
too~\cite{17vb3a,19md2a}. Refs.~\cite{19di3a, 19xc2a,19zw4a} contain broad reviews of
data-driven models used to simulate turbulent flows. Such models fall into three
categories. The first category consists of studies which seek to replace contemporary RANS
closures with new forms learned from TD. This often takes the form of neural
networks~\cite{16lk3a,21zd6a,19zz4a}. Alternatively, studies have used gene expression
programming to assemble new expressions e.g., a linear eddy viscosity model augmented with
additional terms~\cite{16ws2a,20za5a,17ws2a,20sd3a,19sw4a}. The second category consists
of inferring a spatially variable ``correction'' that modulates/multiplies certain closure
terms in the RANS equations; the spatially variable term is then related to the local flow
state with a data-driven model such as a neural net or a random
forest~\cite{16sd2a,16pd2a,17sm3b,19md2a,11dw2a,16ww3a}. This field estimation has
traditionally been performed using optimization, but Kalman filters have also been
used~\cite{16xw5a,17mp2a,18mm1a}.

The third category consists of conventional turbulence closures that have been calibrated
to flows similar to the scenarios where they are expected to be used. Two-equation
turbulence models, like the SST (Shear Stress Transport) model used in this study, contain
a number of approximations~\cite{06wd1a}, which makes it impossible to compute turbulence
closure parameters that are universally generalizable to all types of
flows. Conventionally, these parameters have been calibrated to simple turbulent boundary
layer and shear flows~\cite{96tt2a}, but are routinely tuned to particular classes of
flows~\cite{78ps2a,91sl2a,17sm3c}. Due to the simplicity of the (conventional) closures'
forms, it is usually not possible to estimate the closures' parameters with a great deal
of certainty, and consequently, Bayesian inference is used to construct a JPDF of the
parameters.

The first attempt at Bayesian calibration of closure constants used data from simple flows
e.g., flat-plates and wall-bounded flows~\cite{11co5a,14ec4a}. The 1D flow models used in
these studies did not require surrogate modeling when Bayesian inference was performed
using Markov chain Monte Carlo (MCMC) techniques. In Ref.~\cite{14gg3a} the authors
estimate five closure parameters of the $k-\epsilon$ turbulence model for urban canyon
flows using a Gaussian Process surrogate, MCMC and 10 measurements of turbulent kinetic
energy from a wind-tunnel model. A rather different approach was adopted for the 3D
jet-in-crossflow problem with measurements of velocity and
vorticity~\cite{18rd5a,16rl4a,17rl4a}, where polynomial surrogates were employed to
estimate three closure parameters of the $k-\epsilon$ turbulence model. Bayesian inference
was also used to estimate parameters of the $k-\omega-\gamma$ turbulence model for
hypersonic transitional flows, using Stanton number measurements in a turbulent flow over
a flat plate and polynomial chaos expansion surrogates~\cite{18zf2a}. The SST model has
also been tuned for hypersonic flows, using the same HIFiRE-1 measurements used in this
study~\cite{08mw4a}, but the manual adjustment of a single parameter in the SST model is
not quite a formal calibration. Closure parameters of the $k-\omega$ turbulence model have
also been estimated using Ensemble Transform Kalman Filters and measurements from a
backward-facing step experiment~\cite{16ki3a}. The same method has been used to estimate
spatially-varying turbulent viscosity fields using measurements from transonic flows over
airfoils and wings~\cite{15ky4a}.

\section{The HIFiRE-1  configuration and experiment}
\label{sec:prob}


In this section we describe the flow configuration on which we demonstrate our surrogate
modeling technique.

\begin{figure}[h!]
\centering
\includegraphics[width=10cm]{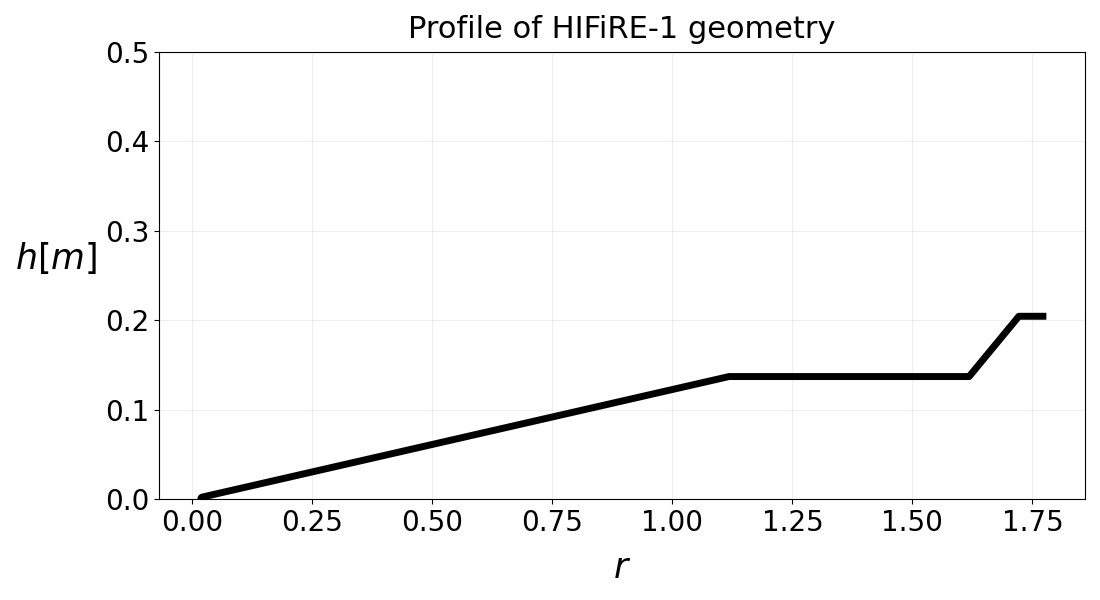}
\caption{Profile of the cone-shaped HIFiRE-1  geometry. The heat flux
and pressure fields are measured as a function of the distance from
the tip of the nose along the axis of rotation, $r$.}
\label{fig: hifire profile}
\end{figure}

\paragraph{The flow configuration: } The flow configuration being simulated is the HIFiRE
ground test conducted in CUBRC's (Calspan-University at Buffalo Research Center) LENS-I
shock-tunnel facility, as described in Ref.~\cite{08wm4a}. The HIFiRE-1 geometry is
cylindrical, and 1721.7 mm in overall length and 409.2 mm in diameter. It consists of a
conical forebody, with half-angle of 7 degrees and of length 1118 mm. The cone has a
blunted nose of diameter 2.5 mm and is followed by a cylindrical midbody of 400 mm. The
aftbody, which is a flare of angle 33 degrees, follows the midbody and is 203.7 mm long
(see schematic of the profile in Fig.~\ref{fig: hifire profile}). The test-section of the
wind-tunnel is capable of accommodating test models 3 feet in diameter and 12 feet
long. The geometry is aligned with the flow, leading to a nominally axisymmetric
flowfield. The HIFiRE-1 surface was instrumented with pressure and heat flux
sensors. There were 42 piezoelectric pressure sensors (with a measurement error of
$\pm 3\%$) and 76 thin-film heat-flux sensors (with a measurement error of $\pm 5\%$) on
the surface. The flow is tripped (artificially rendered turbulent) at a location 0.505
metres from the nose-tip, causing a dramatic increase in aerodynamic heating. The
experiment has been modeled using the RANS equations with the Menter SST turbulence
closure~\cite{94mf1a} (our ``full-order'' model) previously and is described in
Ref.~\cite{08mw4a}. The flow configuration used in this paper is the ``Condition B'' of
Ref.~\cite{08mw4a}, corresponding to nominal HIFiRE-1 flight conditions at an altitude of
21.1 km.  The inflow velocity $v$ is 2170 m/s, with a freestream temperature $u$ of 226.46
K and density $\rho$ of 0.066958 ${\rm kg/m^3}$. The Mach number is 7.16 and the unit
Reynolds number $Re \approx 10.2 \times 10^6 {\rm / m}$. The total enthalpy of the flow is
2.38 MJ/kg. The uncertainty in the LENS-I freestream measurements are 0.5\% for the Mach
number, 3\% for the temperature and 1\% for the pressure~\cite{08wm4a}. Some simple
algebra on the ideal gas model reveals that the freestream velocity and density have a
measurement uncertainty of 2\% each.  The HIFiRE-1 body was kept at a temperature of 296.7
K. The slender cone causes oblique shock-waves to form near the nose.
The turbulent flow separates in front of the flared aftbody, which also causes complex
shock structures, including shock-boundary layer interactions. This is captured in both
the heat-flux and pressure measurements and the full-order model calculations.

\paragraph{The full-order (RANS) model:}
The full-order flow model solves the Reynolds-Averaged Navier Stokes (RANS)
equations~\cite{06wd1a}, with Menter's SST turbulence closure~\cite{94mf1a}. It is
implemented within Sandia's SPARC (Sandia Parallel Aerodynamics and Reentry Code) flow
simulator. SPARC implements a second-order-accurate finite-volume spatial discretization of
the compressible-flow RANS equations, which consists of the continuum conservations laws
for mass, momentum and energy.  These are formulated for reacting gases in thermochemical
non-equilibrium, though for the low total enthalpy of our flow (2.38 MJ/kg) an ideal gas
approximation is used. The equations are solved using a finite-volume method for the
conserved variables. SPARC can accomodate structured and unstructured meshes, though,
given the simple geometry, we only use structured ones in this paper.  For the simulations
in this paper, we use a Steger--Warming scheme for the inviscid fluxes, extended to
second-order using a MUSCL reconstruction. A mimod limiter is used within the
reconstruction. Diffusion and viscous terms in the conservation laws are discretized using
a central difference scheme. SPARC solves the unsteady form of the governing equations,
using a second-order backward difference scheme for the time-integrator. Since all the
simulations in this paper are steady, we run the time-integrator in its first-order form
to accelerate the convergence to a steady-state solution. The HIFiRE-1  simulations were
computed on a 512 $\times$ 256 mesh, clustered near the HIFiRE-1  surface to resolve the
boundary layer and around the region with the shock attached to the nose-tip. Studies to
assess the correct grid resolution were conducted using a sequence of meshes obtained by
doubling the grid resolution along each axis. The finest mesh tested was 4092 $\times$
2048.

\subsection{Model calibration and setup}
\label{sec:model calibration and setup}


The full-order solution represents the heating and pressure fields on the HIFiRE-1 
geometry which we denote by
\begin{equation}\label{eqn:soln}
\mathbf{y}(\cdot;\mathbf{x}) \in \mathbb{R}^{N_s}
\end{equation}
where $N_s$ is the dimensionality of spatial discretization of the solution field, and
$\mathbf{x} \in \mathbb{R}^d$ is the $d$-dimensional tuning parameter or feature space.
\cref{fig: hifire profile} shows the profile of the HIFiRE-1  geometry.  The full geometry
is the shape generated by the rotation of the profile around the $r$ axis. The heat flux
and pressure fields are measured on the surface along the length of the test geometry,
and, due to the symmetry of the mesh, are given as a one-dimensional function of the axis
of rotation, $r$. Plots of the pressure and heat flux fields, computed with the nominal
turbulence model and inflow conditions, can be found in Ref.~\cite{08mw4a}, along with
numerical Schlierens of the shock structures.  We can write the solution vector in
discretized form as
\begin{equation}
\mathbf{y}(\cdot;\mathbf{x}) \doteq [y(r_1;\mathbf{x}),y(r_2;\mathbf{x}),\dots,y(r_{N_s};\mathbf{x})]^T,
\label{eq:soln vector}
\end{equation}
where $r_i$'s are the discretized mesh points along the profile in \cref{fig: hifire profile}.

The feature space consists of 12 tunable input parameters, including three free-stream
parameters, i.e., temperature, density and velocity, and 9 closure constants defined by
the standard Menter two-equation (SST) model~\cite{94mf1a}. These quantities are varied by
scaling (multiplying) them by a uniform random variable resulting in values $\pm 15\%$
from the nominal. \cref{Tab:fs params} shows the three freestream quantities and their
scaling parameters and \cref{tab:sst constants and ranges} does the same for nine SST
parameter constants. The $\pm 15\%$ variation for the freestream quantities' scalings are
designed to bracket the uncertainty in the measured values (see above), so that the
surrogate model is comfortably applicable over our prior belief regarding the freestream
conditions. The uncertainties in the SST parameters are obtained from
Ref.~\cite{17sh6a}. Since these uncertainties are only known as bounds, we proceed with
uniform distributions (under a maximum entropy assumption) for their prior densities.

Note that we are primarily interested in the nine calibrated SST model
parameters. However, the freestream density and velocity in the wind-tunnel, which are
only known to within $\pm 2\%$ error, also affect the measured heat flux and pressure; the
heat flux $q \sim \rho v^3$ and pressure $p \sim \rho v^2$~\cite{20rk10a}. This strong
dependence implies that the uncertainty in the freestream quantities have the potential to
affect the calibrated values of the SST parameters. Thus we will perform a joint
estimation of the freestream variables \emph{and} the SST parameters, and compare it with
a calibration when the freestream parameters held constant at their nominal values. This
comparison will reveal the degree to which the uncertainty in the wind-tunnel inlet conditions
affect the calibrated model.

\begin{table}[h!]
\centering
\begin{tabular}{c|l|c|l}
\hline
\multicolumn{1}{l|}{\textbf{Scaling/multiplier}} & \textbf{Value}  & \multicolumn{1}{l|}{\textbf{Freestream quantity}}  & \textbf{Nominal value}  \\ \hline
$\rho_s$        & $(0.85, 1.15)$ & \cellcolor[HTML]{EFEFEF}$\rho$ (density)   & \cellcolor[HTML]{EFEFEF}$0.066958 $ \\ \hline
$v_s$             & $(0.85, 1.15)$ & \cellcolor[HTML]{EFEFEF}$v$ (velocity)        & \cellcolor[HTML]{EFEFEF}$2170 $        \\ \hline
$u_s$             & $(0.85, 1.15)$ & \cellcolor[HTML]{EFEFEF}$u$ (temperature) & \cellcolor[HTML]{EFEFEF}$226.46 $      \\ \hline
\end{tabular}
\caption{Three parameters (left in white) are the multipliers used to
  scale the freestream specification. The bounds of the uniform
  distribution for them are in the second column. The multipliers are varied by $\pm 15\%$ around 1.
  The third column contains the freestream quantities and the fourth column their nominal values.}
  \label{Tab:fs params}
\end{table}

\begin{table}[h!]
\centering
\begin{tabular}{l|l|l|l|l|l|ll}
\hline
Par. & Value   & Par.  & Value    & Par.   & Value    & \multicolumn{1}{l|}{\cellcolor[HTML]{EFEFEF}Par.}    &  \cellcolor[HTML]{EFEFEF}Value                 \\ \hline
$\sigma_{k_1}$ & $(0.7, 1.0)$ & $\sigma_{w_2}$ & $(0.7, 1.0)$     & $a_1$         & $(0.31, 0.40)$ & \multicolumn{1}{l|}{\cellcolor[HTML]{EFEFEF}$\beta_1$} & \cellcolor[HTML]{EFEFEF}$\beta^*/\beta_{1,r}$ \\ \hline
$\sigma_{k_2}$ & $(0.8, 1.2)$ & $\beta^*$      & $(0.784, .1024)$ & $\beta_{1,r}$ & $(1.19, 1.31)$ & \multicolumn{1}{l|}{\cellcolor[HTML]{EFEFEF}$\beta_2$} & \cellcolor[HTML]{EFEFEF}$\beta^*\beta_{2,r}$  \\ \hline
$\sigma_{w_1}$ & $(0.3, 0.7)$ & $\kappa$       & $(0.38, 0.42)$   & $\beta_{2,r}$ & $(1.05, 1.45)$ &                                                        &                                               \\ \cline{1-6}
\end{tabular}
\caption{Table showing nine parameters from the SST turbulence model
  and their respective parameter ranges. ``Par.'' is an abbreviation
  for SST parameters.}
\label{tab:sst constants and ranges}
\end{table}

\begin{figure}[h!]
\centering
\includegraphics[width=10cm]{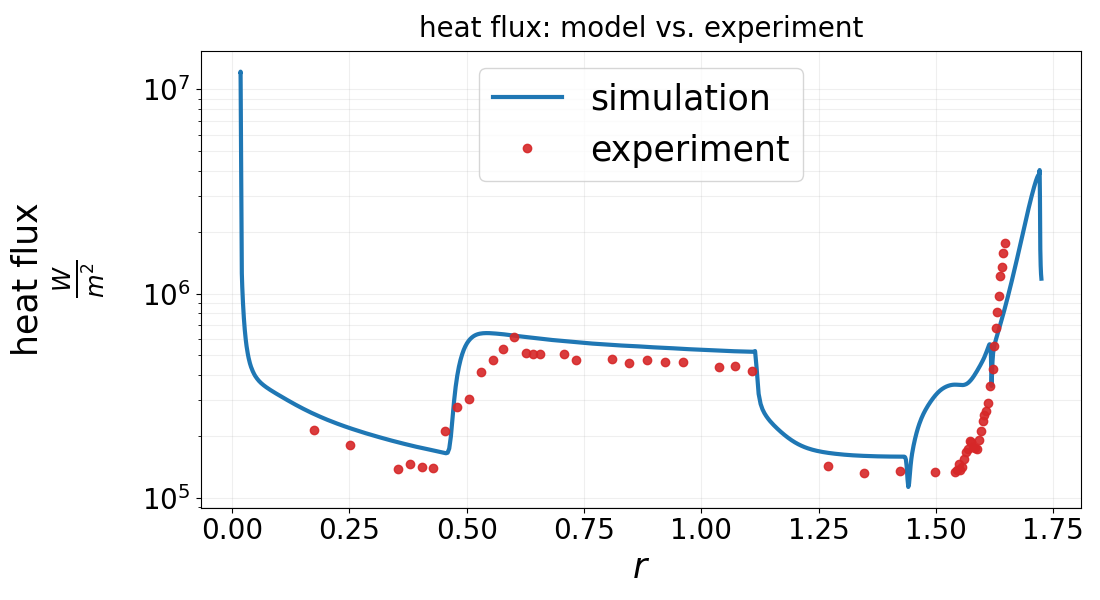}
\includegraphics[width=10cm]{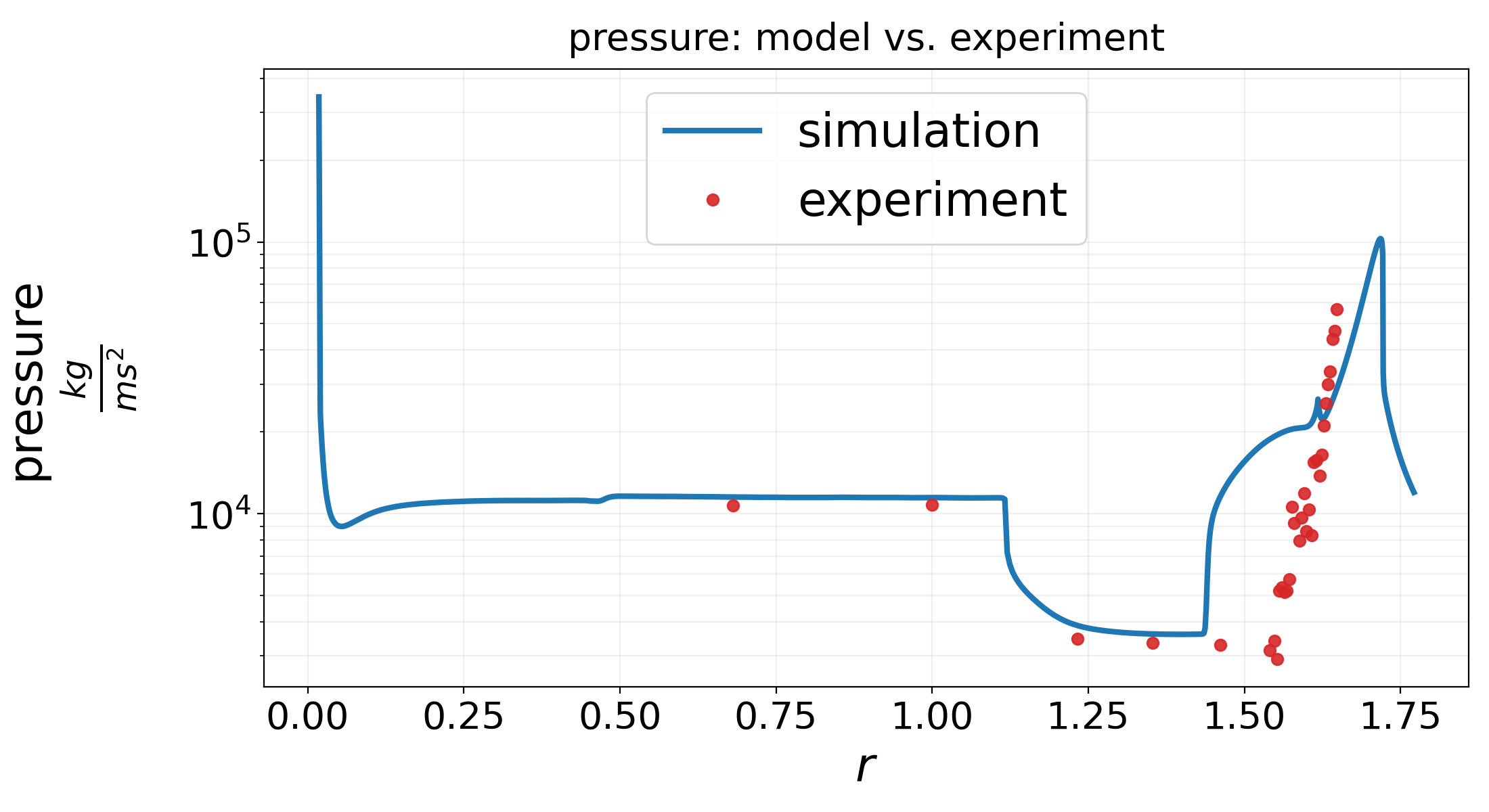}
\caption{Model output versus the experimental HIFiRE-1  data for heat flux and
  pressure. The sharp changes in values correspond to changes in the geometry profile as
  seen in \cref{fig: hifire profile}. The RANS model prediction, using the nominal values
  of the model, is plotted with a solid line. The experimental measurements are plotted
  with symbols.}
\label{fig: model vs exp}
\end{figure}

The goal of calibration is to find the inputs for the flow model (set of $\mathbf{x}$'s)
that minimize some measure of discrepancy between the model prediction, $\mathbf{y}$, and
some observed, experimental data, $\mathbf{y}_{\text{obs}}$. Let
$\mathsf{d}:\mathbb{R}^{n_{\text{obs}}} \times \mathbb{R}^{n_{\text{obs}}} \mapsto
\mathbb{R}^+$ be a discrepancy function between two vectors of size $n_{\text{obs}}$,
which we use to measure the distance between the model prediction
$\mathbf{y}(\cdot,\mathbf{x}) $ and some observation
$\mathbf{y}_{\text{obs}}$.\footnote{We note that $n_\text{obs}$ does not need to be the
  same as, or even a subset of $N_s$. In fact, in this work, the observed points fall in
  between the discretized mesh points.} Then, the deterministic calibration problem can be
written as
\begin{equation}
  \label{eqn:deterministic calibration problem}
  \argmin_{\mathbf{x}} \, \mathsf{d}(\mathbf{y}(\cdot,\mathbf{x}),\mathbf{y}_{\text{obs}},;\theta),
\end{equation}
were $\theta$ represents the parameters of the discrepancy function. A typical discrepancy is the squared error metric given by
\begin{equation} \mathsf{d}_{\text{SE}}(\mathbf{y}(\cdot,\mathbf{x}),\mathbf{y}_{\text{obs}},;\theta) \doteq
   \frac{ \left \lVert \mathbf{y}(\cdot,\mathbf{x}) - \mathbf{y}_{\text{obs}} \right\rVert_2^2 }{\theta^2}
\end{equation}
where the numerator is the canonical squared error norm and $\theta^2$ the variance. In a Bayesian
formulation, we can write
\begin{equation}
  \mathbf{y}_{\text{obs}} = \mathbf{y}(\cdot,\mathbf{x}) + \epsilon,
\label{eq:noise}
\end{equation}
where $\epsilon$ is the discrepancy between model predictions and measurements and is
modeled as $\epsilon \sim \mathcal{N}(0, \theta^2)$, $\mathcal{N}(\cdot, \cdot)$ being a
normal distribution. In such a case, problem~\cref{eqn:deterministic calibration problem}
can be re-interpreted as the negative log-likelihood function. Moreover, if we place a
prior distribution on $\theta$ and $\mathbf{x}$, we have fully defined a posterior
distribution for the feature space parameters:
\begin{equation}
  \label{eqn:bayesian calibration formula}
  \log p(\mathbf{x},\theta) \propto -\mathsf{d}(\mathbf{y}(\cdot,\mathbf{x}),\mathbf{y}_{\text{obs}},;\theta) + \log\pi(\theta,\mathbf{x}),
\end{equation}
where $\pi(\theta,\mathbf{x})$ is the prior distribution on $\theta$ and $\mathbf{x}$,
e.g., an inverse gamma density if $\theta$ represents the variance in a sum of squares
discrepancy error and a uniform prior over some prescribed bounds,
respectively. \cref{fig: model vs exp} illustrates the discrepancy between the full model
prediction, $y(r;\mathbf{x}_i)$ evaluated at a single set of sample parameters and the
experimental data for heat flux and pressure, respectively.
The goal for calibration is to find the values of $\mathbf{x}$ that result in the
\textit{best} match between the model (blue lines in \cref{fig: model vs exp}) and the
experimental data (red dots in \cref{fig: model vs exp}). The search for the optimal
$\mathbf{x}$ will require the evaluation of the RANS model repeatedly, which, at 384
CPU-hours a run, would make the search in 12-dimensional space intractable. Consequently,
it is necessary to replace the RANS model of the HIFiRE-1 geometry with a fast-running proxy
i.e., surrogate model. \textbf{In summary, the calibration procedure involves two stages: (1)
construct a ROM-based surrogate for the heat flux and pressure fields, and then (2) perform Bayesian inference to infer a joint density on the tuning parameters informed by the
discrepancy between the surrogate and the HIFiRE-1 measurements.}


\section{Surrogate construction using proper orthogonal decomposition}
\label{sec:surrogate}

A simple approach for constructing surrogates for spatially varying fields, sometimes called the multi-target regression problem in the
machine learning literature, is to construct a (sub) surrogate for each element in the
output field (see Ref.~\cite{16rl4a} for an example). The complete surrogate is then the
union of all the individual surrogate models. This is cumbersome if the dimensionality of the output 
$\mathbf{y}(\cdot;\mathbf{x})$ is very large, as is the case for most complex problems. In addition, this
method does not preserve the correlations that exist between different
$y(r_i;\mathbf{x})$ in an efficient way. In contrast, we propose using proper orthogonal
decomposition (POD) to transform the solution space to a low-dimensional subspace or
latent space and \textit{then} fit the handful of latent space dimensions with separate
single-target surrogate models. If the transform is invertible\footnote{It may be the case that the inversion is not lossless, e.g., principal component analysis.}, we can
simply invert back to the full order solution space for direct comparison.

The procedure for dimension reduction of the output and subsequent regression fitting is as follows. Let
$\mathbf{Y}\doteq [\mathbf{y}_1,\dots,\mathbf{y}_m]^T \in \mathbb{R}^{m \times N_s}$ be a
snapshot matrix of sample solutions or ensemble runs, where each row of $\mathbf{Y}$ represents a solution field for a particular parameter set. These $m$ samples are generated by sampling the
twelve-dimensional feature space using Latin hypercube sampling over the prescribed
bounds. We then perform principal component analysis (PCA) on this (centered) snapshot matrix to
obtain a set of orthogonal transformations denoted by
$\mathbf{\Phi}=[\bm{\phi}_1,\dots,\bm{\phi}_{\rbDim}] \in \mathbb{R}^{N_s \times \rbDim}$,
where $\bm{\phi}_i \in \mathbb{R}^{N_s}$'s represent $n$ new coordinate axes representing the directions of maximum variances. The associated coordinates or projection coefficients 
for each basis term is given by $\mathbf{c}_i \doteq \mathbf{Y}_c \bm{\phi}_i \in \mathbb{R}^m$, where $\mathbf{Y}_c$ is the centered snapshop matrix (see \cref{alg:PCA}). The empirical
variance is then given by $\lambda_i \doteq \sigma_i^2/(m-1)$, where $\sigma_i$'s are the singular values
associated with the SVD of $\mathbf{Y}$, or equivalently, the eigenvalues associated with
the normal matrix.
With $\mathbf{X}\doteq [\mathbf{x}_1,\dots,\mathbf{x}_m]^T \in \mathbb{R}^{m \times d}$ as
the data matrix for the feature space, our subsequent task is to then create a surrogate
model for each of the reduced space training data pairs $\{\mathbf{X}_i,\mathbf{c}_i\}$ for
$i = 1, \dots,n$. While this is still a multi-target regression problem, the number of targets is $n \ll N_s$. If we denote
$\hat{y}_j(\mathbf{x}):\mathbb{R}^d \mapsto \mathbb{R}$ each of the $j=1,\dots,n$
surrogate models corresponding to each component, then our full surrogate model is given
by the Karhunen-Loeve expansion~\cite{16ch2a} 

\begin{equation}
\mathbf{y}(\cdot;\mathbf{x}) \approx \tilde{\mathbf{y}}(\cdot;\mathbf{x}) = \mathbf{\mu}_0 +\sum_{j=1}^{\rbDim} \sqrt{\lambda_j }\hat{y}_j(\mathbf{x})\bm{\phi}_j.
\label{eq:linear expansion} 
\end{equation}
The complete PCA/ POD algorithm with details about automating the choice of $n$ is shown in
\cref{alg:PCA}.
\begin{center}
\begin{minipage}[t]{16cm}
\vspace{0pt}
\begin{algorithm}[H]
\begin{algorithmic}[1]
  \caption{Principal Component Analysis}
  \label{alg:PCA}
  \REQUIRE Snapshots $\mathbf{Y} \in \mathbb{R}^{m \times N_s}$ and percentage variance threshold $\nu \in [0,1]$.\footnote{Each of the $m$ snapshots corresponds to a model output evaluated at $\mathbf{x}_m$, i.e. a sample from the feature or tuning space.}
  \ENSURE Basis matrix $\mathbf{\Phi} \in \mathbb{R}^{N_s \times k^*}$, and projections $\mathbf{C} \in \mathbb{R}^{m \times k^*}$. 
  \STATE Center the snapshot data matrix, i.e. $\mathbf{Y}_c = \mathbf{Y} - \bm{\mu}$, where $\bm{\mu} = \sum_{j=1}^m \mathbf{y}_j$, i.e., mean w.r.t. the rows. 
  \STATE Compute thin singular value decomposition (SVD): $\mathbf{Y}_c = \mathbf{U} \mathbf{\Sigma} \mathbf{V}^T$, where $\mathbf{U} \in \mathbb{R}^{m \times K}$, $\mathbf{\Sigma} \in \mathbb{R}^{K \times K}$, $\mathbf{V} \in \mathbb{R}^{N_s \times K}$, where $K = \text{min}(N_s,m)$.
  \STATE Find $k^* = \argmin\{k \in \mathbb{N}^+:\sum_i^k \sigma_i^2 /\sum_i^K \sigma^2_{i} \geq \nu\}$. 
  \STATE Set $\mathbf{\Phi} = [\mathbf{v}_1,\dots,\mathbf{v}_{k^*}]$, where $\mathbf{v}_i$'s are the columns of $\mathbf{V}$ and compute $\mathbf{C} = \mathbf{Y}\mathbf{\Phi}$.\footnote{One can also scale the projections using $\mathbf{C} = \mathbf{Y}\mathbf{\Phi}\mathbf{\Sigma}_*^{-1}$, where $\mathbf{\Sigma}_*$ is the $k^* \times k^*$ submatrix of $\mathbf{\Sigma}$.}
\end{algorithmic}
\end{algorithm}
\end{minipage}
\end{center} 

Once the PCA approach is performed on the solution field, the remaining effort is constructing machine learning surrogates for $\hat{y}(\mathbf{x})_j$'s. To determine the best regression, we experiment with an array of different types of machine learning regressor models
including Gaussian process regression, multi-layer perceptron (fully connected neural network) models, random forests,
kernel ridge regression, support vector machines, and, last but not least, polynomial
chaos (or multi-variate polynomial) expansions using Legendre polynomials. Each of these
regressors are hyper-parameter-tuned over a specified parameter grid, e.g., polynomial
order and regularization type for polynomial fitting, using five-fold cross validation in
order to perform model comparison (see \cref{appendix: rom models}) . We briefly summarize some of the key model features of
these model regressors in Sec.~\ref{sec:rom}. See Ref.~\cite{05xi2a,hoang2021projection} for a
more thorough discussion of orthogonal polynomial interpolants for multivariate model
fitting and dense neural network construction, and the Scikit-Learn
documentation~\cite{scikit-learn} for a brief discussion of the other five estimators and
their respective implementation.

\section{Surrogate models for HIFiRE-1 simulations }
\label{sec:rom}

\begin{figure}[h!]
\centering
\includegraphics[width=7.5cm]{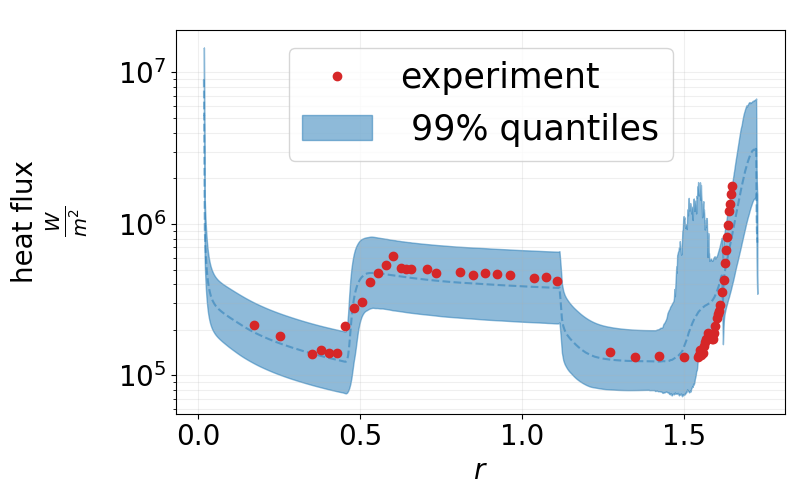}
\includegraphics[width=7.5cm]{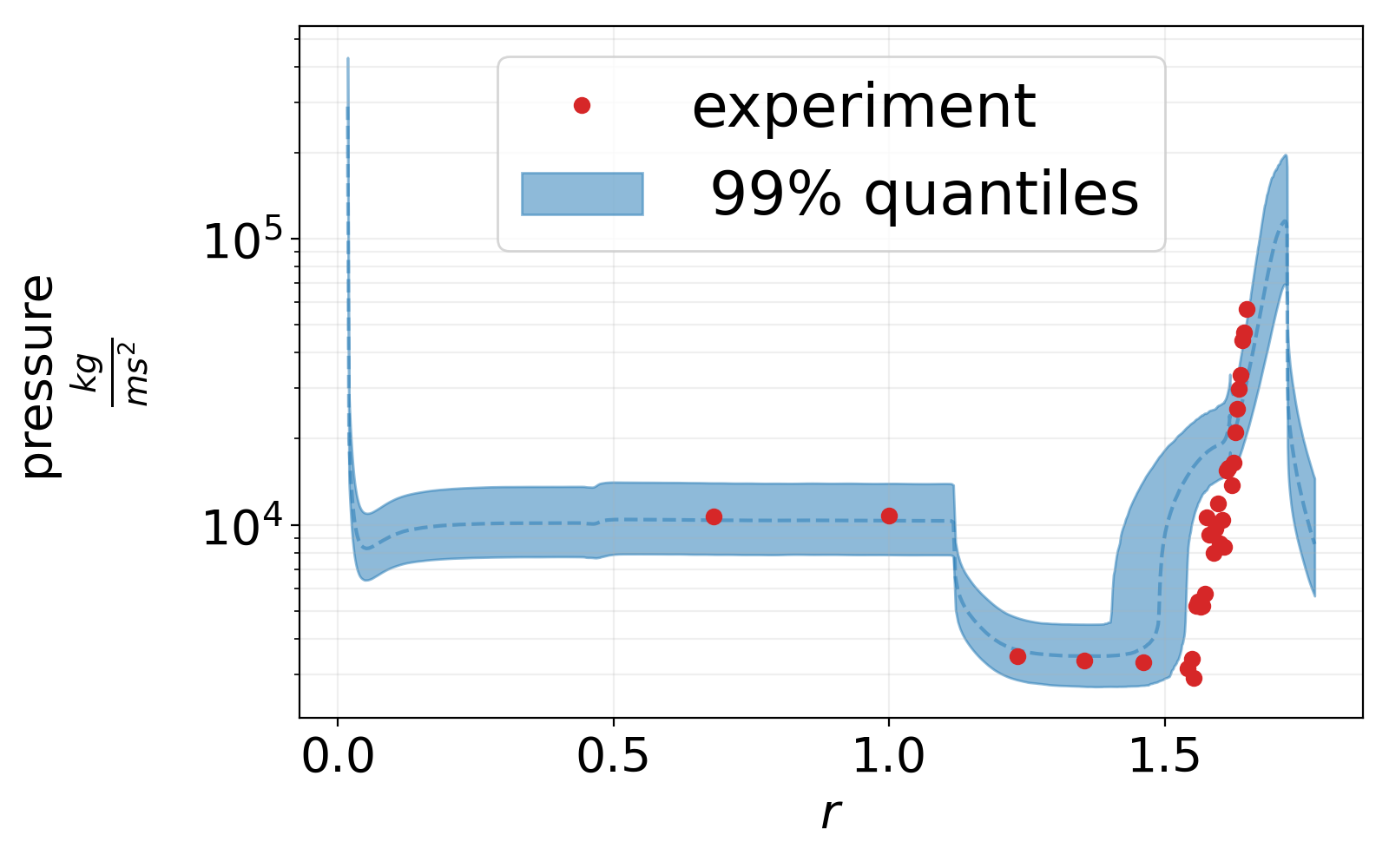}
\caption{1\% quantiles (for each $r_i$) of heat flux and pressure fields from LHS study,
  simulated at $m = 2,500$ LHS points, plotted against the geometric profile in \cref{fig:
    hifire profile}. The experimental measurements for heat flux and pressure,
  respectively, are also shown in red. Note that for $r \lesssim 1.5$, the 99\% quantiles
  of the LHS study encompasses the experimental data, but beyond that the numerical
  simulations struggle to capture the effects. This is most likely due to the limitations
  (model-form error) of RANS. The median is shown in the dashed blue line.}
\label{fig: LHS study}
\end{figure}

The training dataset (TD) for the calibration and surrogate model construction is generated using Latin hypercube
sampling (LHS) of the feature space.Ranges for the LHS study are defined in ~\cref{Tab:fs
  params} and \cref{tab:sst constants and ranges}.\footnote{Note that the ranges for the training data are larger than the allowable ranges for the Bayesian calibration in order to avoid problems with extrapolation.}  The full-order i.e., RANS model is then
evaluated at $m = 2,500$ sample points and the heat flux and pressure fields are recorded
to produce input/output data pairs $\{(\mathbf{X}_i,\mathbf{y}_i)\}$ for
$i = 1,\dots,n$. Furthermore, we use 5-fold cross validation to tune and evaluate the accuracy of each possible regression technique (see \cref{appendix: rom models} for more details about the hyper-parameter tuning). \cref{fig: LHS study} shows summary statistics of realizations or\footnote{Note that the ranges for the training data are larger than the allowable ranges for the Bayesian calibration in order to avoid problems with extrapolation.} 
snapshots of the heat flux and pressure fields. The realizations are plotted as a function
of the geometry profile shown in \cref{fig: hifire profile}. This is data we use to train
our surrogate model in order to capture the effect of perturbations in the
twelve-dimension feature space on the heat flux and pressure fields.

\begin{figure}[h!]
\centering
\includegraphics[width=7.5cm]{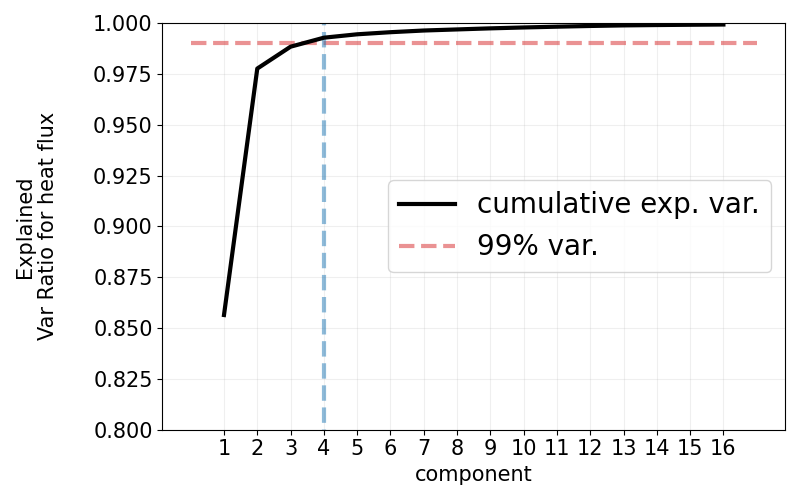}
\includegraphics[width=7.5cm]{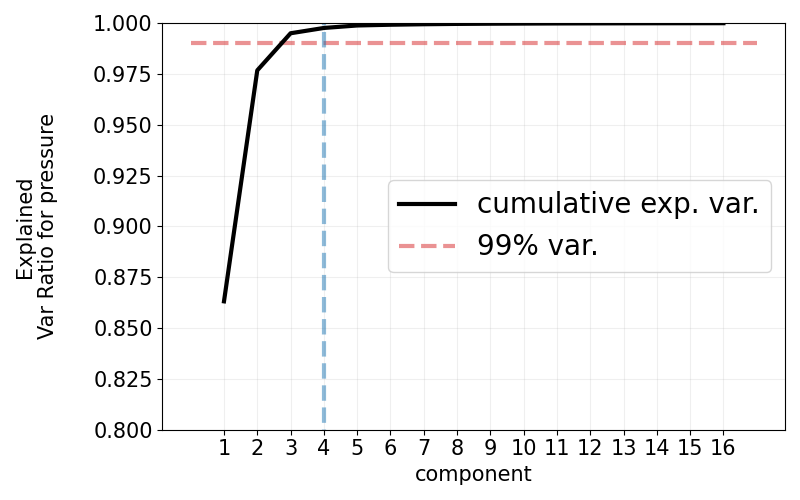}
\caption{Cumulative explained variance ratio (i.e., Scree plots) as a function of the PCA component for both
  the heat flux (left) and pressure (right) fields. This is equivalent to the cumulative sum of the square of the singular values in the PCA algorithm (\cref{alg:PCA}). Four components (vertical blue line) are enough to capture
  more than 99\% of the total variance (red dashed horizontal line) for both heat flux and pressure fields. This means that we can reduce our dimensionality almost a thousand fold to four dimensions.}
\label{fig: cum pca exp var ratio}
\end{figure}

Once the TD is generated, we begin with the dimensionality reduction of the spatially
varying targets using \cref{alg:PCA}.  \cref{fig: cum pca exp var ratio} shows the
cumulative explained variance ratio, which can be used to determine the dimension of the
latent space. From this plot we can deduce that a latent space dimension of $\rbDim = 4$
captures more than 99\% of the total variance of the original signal. \footnote{We
  experimented with using six or eight components, but the resulting surrogate only
  improved test errors by less than a tenth of a percent.}

\begin{figure}[h!]
\centering
\includegraphics[width=7cm]{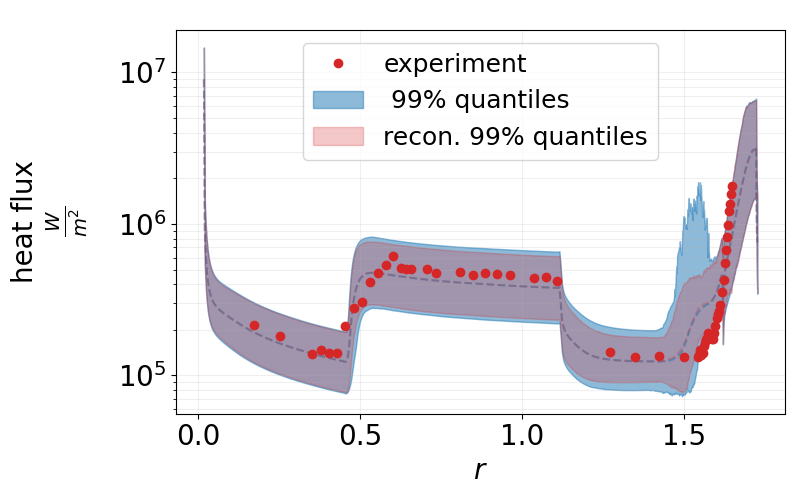}
\includegraphics[width=7cm]{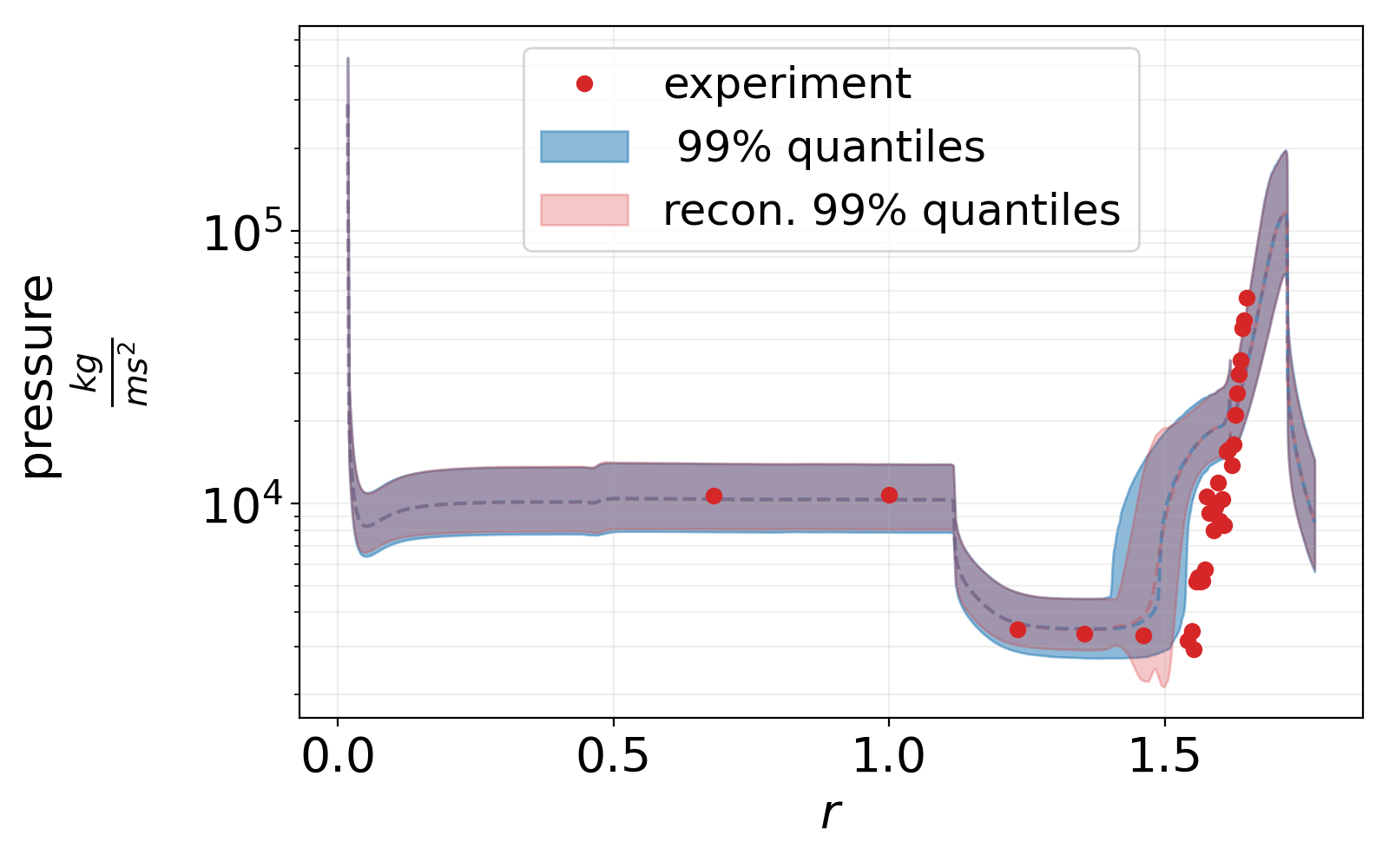}
\caption{99\% quantile comparisons of heat flux and pressure fields from LHS study (blue) versus PCA reconstruction samples (red), compared to experimental data (red dots). }
\label{fig: LHS vs PCA study}
\end{figure}

The first four components for the heat flux and pressure fields are shown in \cref{fig:
  pca components plots}. The components (plotted as a function/vector of the geometric
profile parameter $r$) represent the directions of maximum variance, in decreasing
order. One can also interpret these components or modes as the axes of a new coordinate
transformation in $N_s$-dimensional space.
\begin{figure}[h!]
\centering
\includegraphics[width=7.5cm]{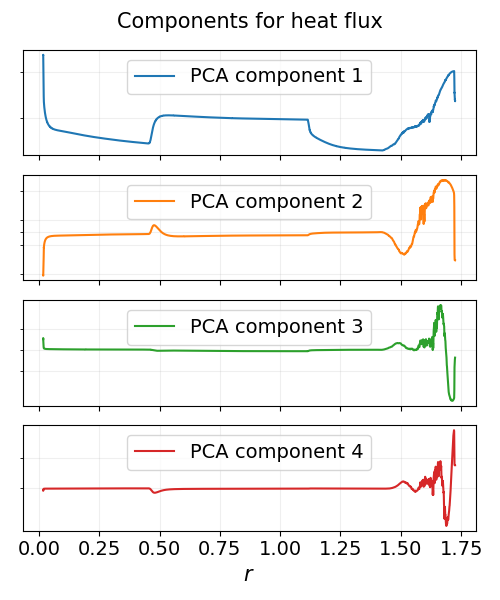}
\includegraphics[width=7.5cm]{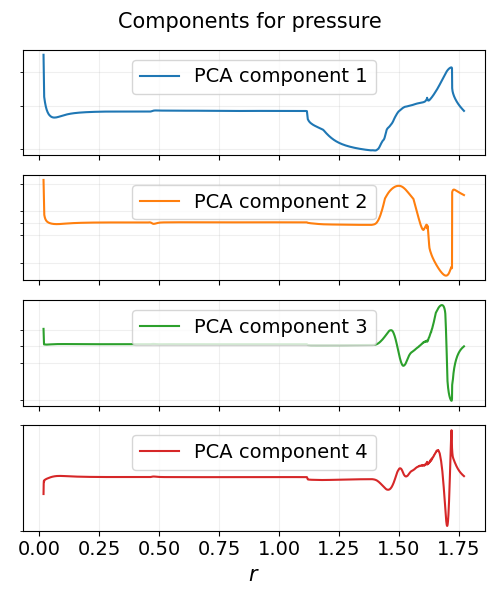}
\caption{First four principal components for heat flux (left) and pressure (right)
  fields. Much of the variance is concentrated around the $r \gtrapprox 1.5$ area, where
  the simulation struggles to capture the experimental data.}
\label{fig: pca components plots}
\end{figure}
While only four components are needed to capture nearly the entirety of the variance,
\cref{fig: LHS vs PCA study} shows the difference in the original versus the reconstructed
signal using $\rbDim=4$ components. Indeed the statistics of reconstructed signal are indeed distinguishable from the
original data. This shows that the dimension reduction, while extremely accurate in
capturing the total variance, is not lossless when transformed back to the original space.

\begin{figure}[h!]
\centering
\includegraphics[width=7.5cm]{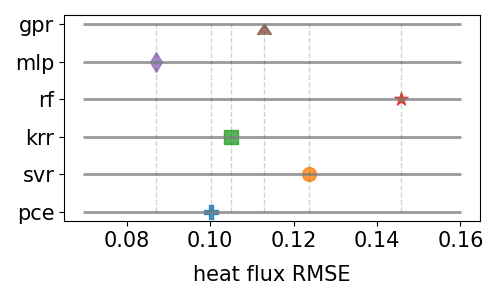}
\includegraphics[width=7.5cm]{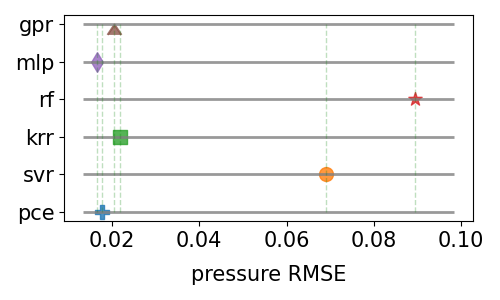}
\caption{Root mean squared errors for regression surrogates for
  the first projection component of heat flux (left) and pressure (right). The six
  different models used for comparison are Gaussian processes (gpr), multi-layer
  perceptron (mlp) or dense neural networks, kernel ridge regressor (krr), random forests
  (rf), support vector machine (svr), and polynomial chaos expansions (pce). }
\label{fig: rmsRe for yhat}
\end{figure}

Each of the component projections are fit with six different regression models and a
5-fold cross-validation score is computed. Furthermore, for added robustness, each
regressor is tuned over a set of prescribed hyper-parameters, e.g., a range of maximum
depth and number of estimator are tested to obtain the best overall random forests
regressor, a range of polynomial orders, least squares and sparse solvers are tested to
obtain the best overall polynomial regressor, and etc. (See \cref{appendix: rom models} for more details about the hyper-parameter tuning and the final model architecture). We use a negative root mean square
score function for the individual regressors.  \cref{fig: rmsRe for yhat} shows the root
mean square errors (RMSE) for $\hat{y}_1$ in \cref{eq:linear expansion}, i.e., the surrogate for the projection coefficients of the first component of heat flux and pressure, respectively. The $y$-axis separates the different regressors and the $x$-axis shows the the RMSE (smaller the better). Recall that
the first component has the largest contribution to the total variance, and it decreases
from there onwards. Thus, if we focus on the first component, we see that the two best
regressors, i.e., the lowest RMSE, for fitting the component projections 
are the polynomial chaos expansions (PCE) with Legendre polynomials and the multi-layer
perceptron (MLP) models. We left out errors for the k-nearest neighbor approach since the errors where significantly worse. 

We also compute the root mean square relative error (RMSRE) for the full field solution
$\tilde{y}(\bm{x})$, which we obtain by projecting the latent space surrogate back to the
original space, for the two best performing surrogates in the latent space, i.e., the
PCEs and the MLPs. The error for the full field solution is defined by
\begin{equation}
\epsilon_i \doteq \displaystyle\frac{\|\mathbf{y}_i-\tilde{\mathbf{y}}_i\|^2}{\|\mathbf{y}_i\|^2},
\end{equation}
where
$\tilde{\mathbf{y}}$ is the ML surrogate, i.e. either PCE or MLP, evaluated at the $i^{th}$ training or observed data point. Here we haven chosen to use relative error explicitly in order to get a sense of the
relative magnitude of the surrogate construction error, and not just as a tool for model
selection. Not shown in this plot is the baseline error computed by creating a ``dummy''
mean predictor which had a RMSRE of .1. Thus, our surrogate model reduces the relative
error by almost an order of magnitude (10x) from the mean predictor.

\begin{figure}[h!]
\centering
\includegraphics[width=7.5cm]{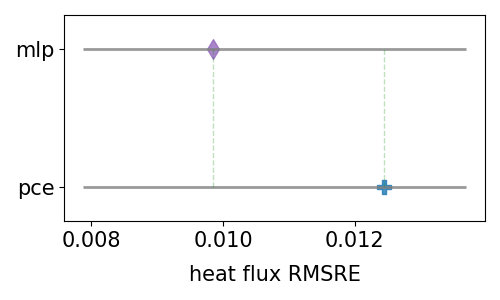}
\includegraphics[width=7.5cm]{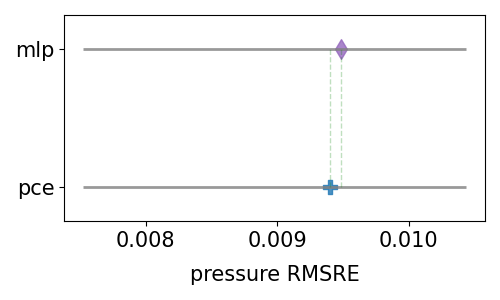}
\caption{Root mean squared relative errors (RMSRE) for the full field solution surrogate. We
  calculate the cross-validation error only for two best performing models on the
  projection coefficients - multi-layer perceptron (mlp) or dense neural networks and
  polynomial chaos expansions (pce). Training of the full ROM-based surrogate model is
  performed on a separate test set, in order to avoid biased error calculations, using
  5-fold cross-validation. Heat flux is shown in the left and pressure on the right. For a
  baseline comparison, the mean ``dummy'' predictor has an RMSRE of .1, which is roughly ten times worse.}
\label{fig: rmsRe for y}
\end{figure}

When choosing our final surrogate model, we have to balance simplicity with accuracy. With
that in mind, for the purposes of the Bayesian calibration, we will use Legendre PCEs for
our surrogate model construction. MLPs suffered from long training times, difficult
hyper-parameter tuning spaces, and sensitivities to random seeds, which made them less
reliable and less robust. PCEs offered simpler models and fitting approaches with far fewer tuning parameters and better reproducability. 

Before we move on the Bayesian calibration we would like to make a quick remark about the speed up in computational time. Recall that a single run for the full simulation takes approximately three hours on a supercomputer. The surrogate, on the other hand, can run in less than a tenth of a second on a standard laptop. Thus, the surrogate runs about one hundred thousand times faster, i.e., a speed up of five orders of magnitude, with far less computational power! The caveat is that, of course, the generation of the training data and the subsequent training of the surrogate model itself are not trivial tasks. Still, without the surrogate, it would be impossible to perform any sort of gradient-based parameter optimization or Bayesian calibration, which requires thousands or even millions of model evaluations, in any reasonable amount of time.  

Next, in Sec.~\ref{sec:calib}, we use the newly constructed ROM-based PCE surrogate to
perform Bayesian calibration of the HIFiRE-1 experiment. We will show that the Bayesian
approach not only provides a better overall model, i.e., improved match to the HIFiRE
experiment, but also a measure of the uncertainty for the heat flux and pressure fields
via the joint parameter probability density function.

\section{Bayesian calibration of the HIFiRE-1 experiment}
\label{sec:calib}

As in \cref{sec:model calibration and setup}, we use a squared error discrepancy term,
which equates to standard additive Gaussian white noise model (Eq.~\ref{eq:noise}; also
Ref.~\cite[Section~2.5]{gelmanbda04}), between the surrogate predictions and the
experimental data. Since we are attempting to simulatenously use both the heat flux and
the pressure field during the calibration, our final discrepancy error is actually a
normalized average of the error in the heat flux and the pressure field, with the same
shared $\mathbf{x}$ parameter, but separate noise models $\theta_q$ and $\theta_p$. This
implies that the mismatch between model predictions and measurements at a point for the
heat flux is assumed to be independent of the mismatch observed for the pressure. Denote
$\mathsf{d}_q(\mathbf{x};\theta_q)$ and $\mathsf{d}_p(\mathbf{x};\theta_p)$ to be the heat
flux and pressure log-likelihood discrepancy terms respectively, where $\theta_p$ and
$\theta_q$ represent the noise parameters of the log-likelihood. Then, the full model
posterior form is given by
\begin{equation}
\log p(\mathbf{x};\theta_q,\theta_p) \propto -w_q\mathsf{d}_{q}(\mathbf{x};\theta_q) - w_p\mathsf{d}_{p}(\mathbf{x};\theta_p) + \log \pi(\theta_q)  + \log \pi(\theta_p) + \log \pi(\mathbf{x}),
\end{equation}
where $\log \pi(\theta_q)$, $\log \pi(\theta_p)$, and $\log \pi(\mathbf{x})$ are the log
priors, and $w_p$,$w_q$ are weights chosen to give equal weighting to pressure and heat
flux calibration.\footnote{Since the number of heat flux, $n_{\text{obs}}^q$, and
  pressure, $n_{\text{obs}}^p$, observations are different, we set
  $w_q=1/n_{\text{obs}}^q$ and $w_p=1/n_{\text{obs}}^p$ or, equivalently, $w_q=1$ and
  $w_p=n_{\text{obs}}^q/n_{\text{obs}}^p$, to give equal weight to all observations.}  The
explicit posterior is given by
\begin{align}
  \label{eq:full posterior}
\log p(\mathbf{x};\theta_q,\theta_p) \propto
  -\frac{w_q}{2\theta_q^2}\sum_{i=1}^{n_\text{obs}^q} \left(\tilde{y}_q(r_i^{\text{Exp}};  \mathbf{x}) - y_{q,i}^{\text{Exp}}\right)^2 - w_q n_\text{obs}^q \log\theta_q \nonumber \\ 
-\frac{w_p}{2\theta_p^2}\sum_{i=1}^{n_\text{obs}^p} \left(\tilde{y}_p(r_i^{\text{Exp}};  \mathbf{x}) - y_{p,i}^{\text{Exp}}\right)^2 - w_p n_\text{obs}^p \log\theta_p \nonumber \\ 
 + \log \pi(\theta_q)  + \log \pi(\theta_p) + \log \pi(\mathbf{x}),
\end{align}
where $n_\text{obs}^q$ is the number of experimental observations for heat flux,
$r_i^{\text{Exp}}$'s are the locations at which the observations were made (different from
$r_i$), $y_{q,i}^{\text{Exp}}$ is the experimental observations for heat flux, and
$\tilde{y}_q(r_i^{\text{Exp}})$ is the surrogate prediction at the observed location. The
terms are analogous for pressure. It is clear from \cref{fig: LHS study} that the
discrepancy between the model and the observations are higher for the heat flux than the
pressure (even after weighting), thus the reason for the two $\theta$'s.

We briefly discuss the choice of the prior distributions next. The feature vector
$\mathbf{x}$ are divided into two sets. The first set, consisting of the (scalings for)
freestream (or wind-tunnel inlet) conditions, are modeled using uniform distributions as
$\rho_s \sim \mathcal{U}(0.98, 1.02)$, $v_s \sim \mathcal{U}(0.98, 1.02)$ and
$u_s \sim \mathcal{U}(0.85, 1.15)$, where $\mathcal{U}(a, b)$ denotes a uniform
distribution between $(a, b)$. The bounds for $\rho_s$ and $v_s$ reflect the $\pm 2\%$
uncertainty in the freestream conditions for density and velocity (see
Sec.~\ref{sec:prob}). The measured quantities are only weakly sensitive to the freestream
temperature and so the bounds on $u_s$ are the same as those used for training the
surrogate model (\cref{Tab:fs params}). The second set consists of the SST model
parameters whose prior densities are cast as uniform distributions with the bounds
specified in \cref{tab:sst constants and ranges}. We use a gamma prior on the inverse variance parameter, which is the conjugate prior for the Gaussian likelihood, where we denote the precision as $\tau_q=\theta_q^{-2}$ and
$\tau_p=\theta_p^{-2}$ with shape and scale parameters for their respective gamma densities set to $k=2,\theta=2$, chosen to
encapsulate the model errors.\footnote{The log gamma prior for $\tau$ is
  $\log(\tau;k,\theta) \propto (k-1)\log(\tau) - \frac{\tau}{\theta}$}

\subsection{MCMC results}
\label{sec:mcmc}

In order to obtain samples from our posterior distribution, we use Markov Chain Monte
Carlo (MCMC)~\cite{96gr3a,10ll3a}. The idea of MCMC is to derive a Markov chain,
with a prescribed transition probability, such that the chain converges to a
stationary distribution equal to the posterior distribution.  A
modified version of the classic Metropolis-Hastings algorithm \cite{gelmanbda04} which
adapts the covariance kernel of the transition probabilities
\cite{Haario2006} was used, often referred to as an adaptive MCMC methods (AMCMC).  32 parallel chains weree run, each with a 50,000 burn-in period and 750,000 post burn-in runs, for
a total of 24 million samples. The resulting chains had an average autocorrelation  of $\leq 500$ and all chains had an acceptance rate of $~0.21 - 0.22$. We
aggressively thinned the chain by 1000 for a total of effective samples size of 24,000
samples.  See Appendix~\ref{sec:bcr} for autocorrelation diagnostics.

\begin{figure}[h!]
\centering
\includegraphics[width=16cm]{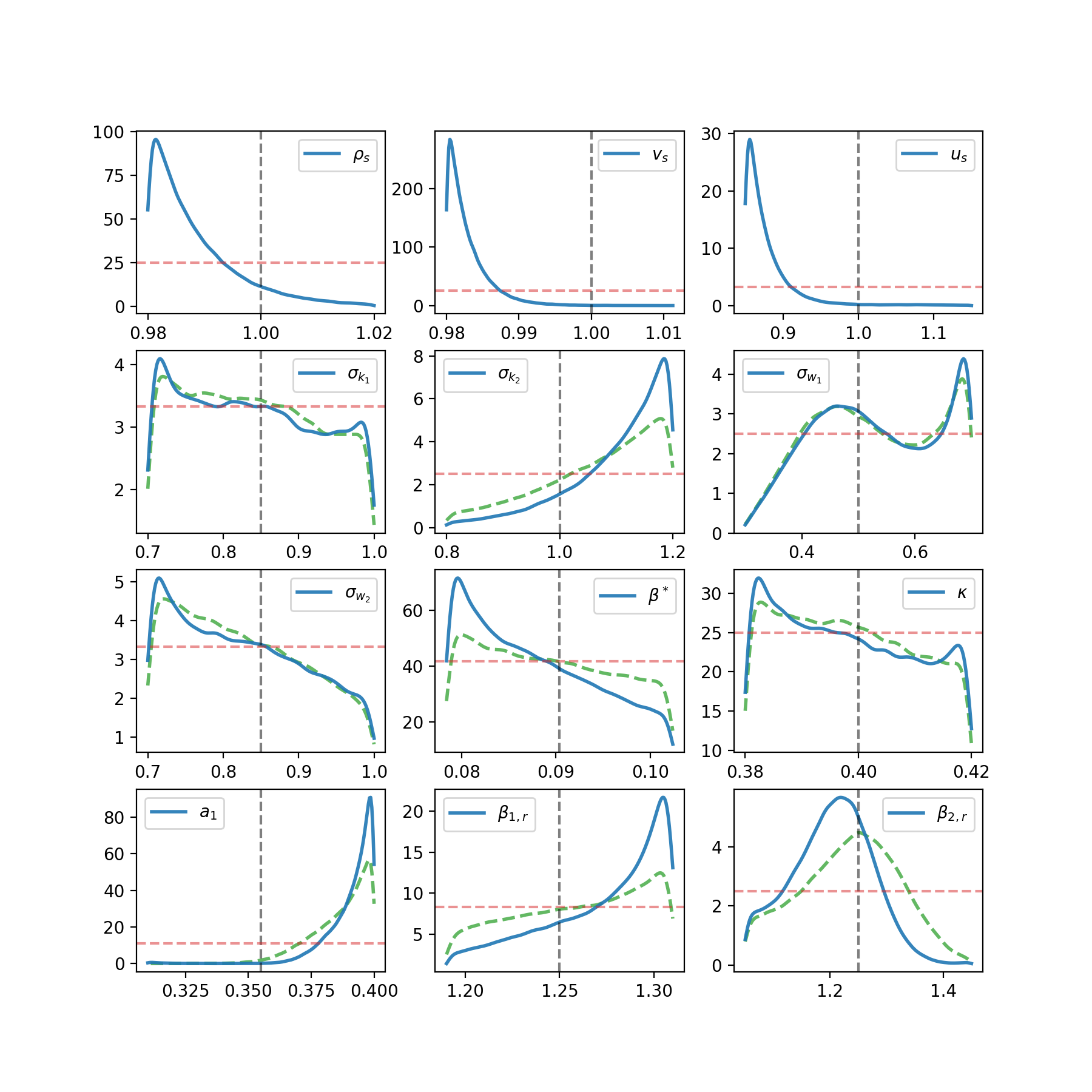}
\caption{Marginalized posterior PDFs for the 12 features, plotted using a solid line. The
  uniform priors are plotted using red dashed lines. The vertical line denotes the nominal
  value of the parameters. The green dashed line plots the PDFs of the SST model parameters
  inferred after assuming that the freestream conditions are perfectly known and are at
  their nominal value. This means that the free stream parameters are inferred independently from the SST constants, which can be observed from the joint density in \cref{fig: posterior corner plot}.}
\label{fig: posterior univariate histograms}
\end{figure}

The thinned chains provide samples from the 12-dimensional JPDF that is the solution of
the Bayesian inverse problem for the freestream scalings and SST parameters. These samples
are marginalized (integrated over all dimensions except one) to compute the posterior
probability density functions (PDFs) of each of the features i.e., elements of
$\mathbf{x}$. These are plotted in \cref{fig: posterior univariate histograms} with a
solid line. The prior distribution (plotted with a red dashed line) and the nominal value
taken from \cref{Tab:fs params} and \cref{tab:sst constants and ranges} (vertical dashed
line), are also shown. A posterior PDF that differs significantly from the prior density
implies a calibrated parameter that has assimilated information from the measurements. It
is clear that the freestream scaling $(\rho_s, v_s, u_s)$ can be inferred quite easily -
the PDFs' peaks are sharp and distinct from the nominal values. The performance of the SST
parameters are mixed. Some like $a_1$ and $\sigma_{k_2}$ have sharp PDFs whereas others
such as $\sigma_{k_1}$ and $K$ are not well informed by the measurements.

\begin{figure}[h!]
\centering
\includegraphics[width=12cm]{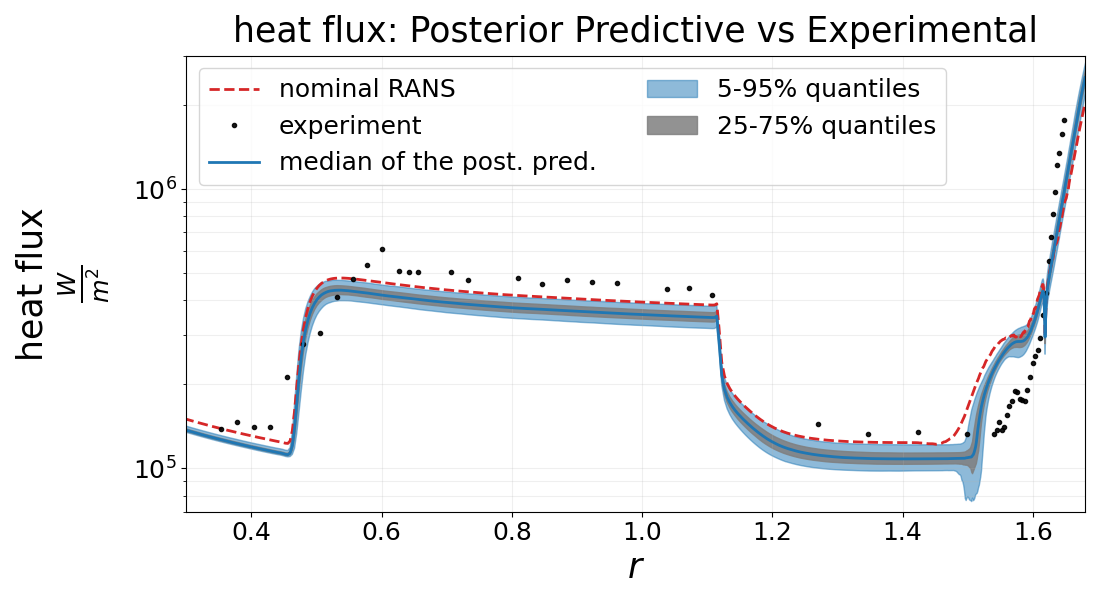}
\caption{Comparison of heat flux predictions before and after calibration. The prediction
  using the nominal SST model is plotted dashed red line, the median prediction with a
  solid blue line, the first and third quartiles with a dashed blue line and the
  ${\rm 5^{th}/95^{th}}$ percentiles with a dotted blue line. The experimental data is
  plotted with symbols. We also plot the median solution of the posterior predictive shown in a dark blue line. }
\label{fig: posterior pred vs exp heat flux}
\end{figure}

\begin{figure}[h!]
\centering
\includegraphics[width=12cm]{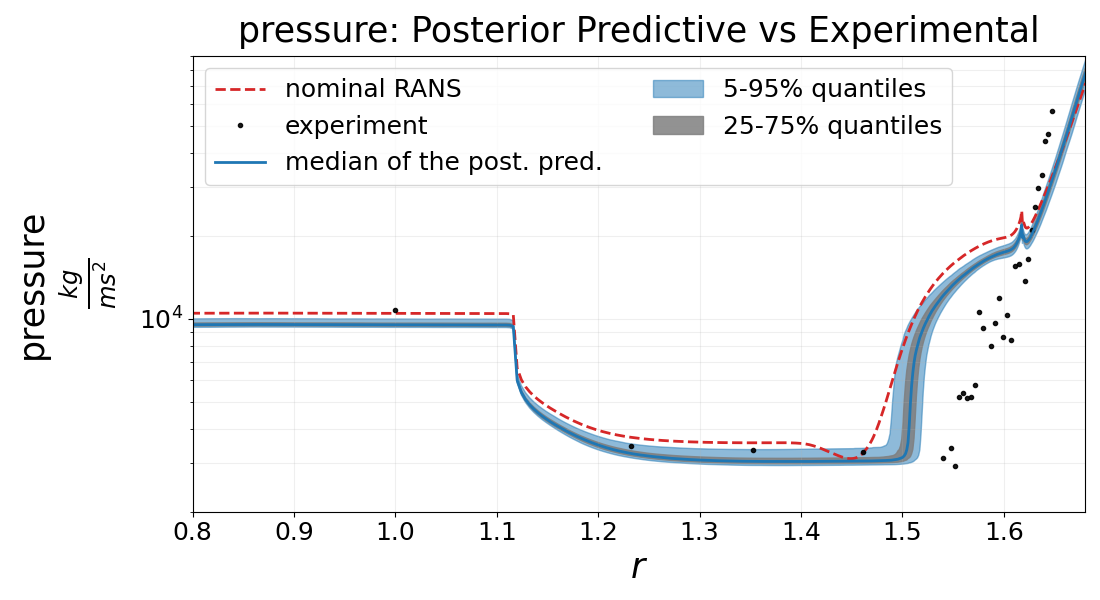}
\caption{Comparison of pressure predictions before and after calibration. The prediction
  using the nominal SST model is plotted dashed red line, the median prediction with a
  solid blue line, the first and third quartiles with a dashed blue line and the
  ${\rm 5^{th}/95^{th}}$ percentiles with a dotted blue line. The experimental data is
  plotted with symbols. We also plot the median solution of the posterior predictive shown in a dark blue line.}
\label{fig: posterior pred vs exp pressure}
\end{figure}

The final analysis involves pushing through a few hundred samples of the posterior back
through the RANS model in order to determine if our Bayesian procedure actually results in
a better calibration. These ``pushed-forward-posterior'' simulations lead to a
distribution of predicted heat fluxes and pressures, which are summarized in \cref{fig:
  posterior pred vs exp heat flux} for the heat flux predictions and \cref{fig: posterior
  pred vs exp pressure} for the pressure. The prediction using the nominal SST model is
plotted with a dashed red line, the median prediction with a solid blue line, the first
and third quartiles with a dashed blue line and the ${\rm 5^{th}/95^{th}}$ percentiles
with a dotted blue line. The experimental data is plotted with symbols. We see that in the
region over the cone with turbulent flow ($0.4 \leq r \leq 1.1$), calibration reduces the
agreement with heat flux measurements (\cref{fig: posterior pred vs exp heat flux}) though
the agreement improves in the highly instrumented separation zone at the back of the
HIFiRE-1 geometry ($1.5 \leq r \leq 1.6$). A similar effect, though much more muted, occurs
with pressure predictions (\cref{fig: posterior pred vs exp pressure}).

\begin{figure}[h!]
\centering
\includegraphics[width=12cm]{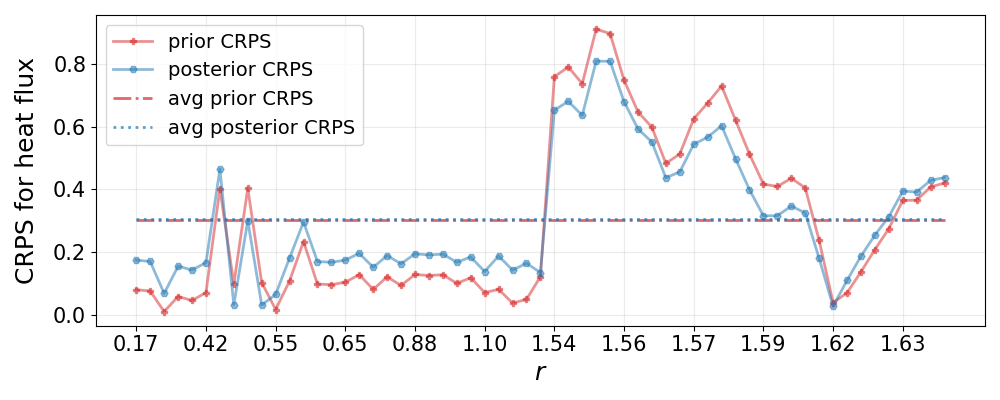}
\includegraphics[width=12cm]{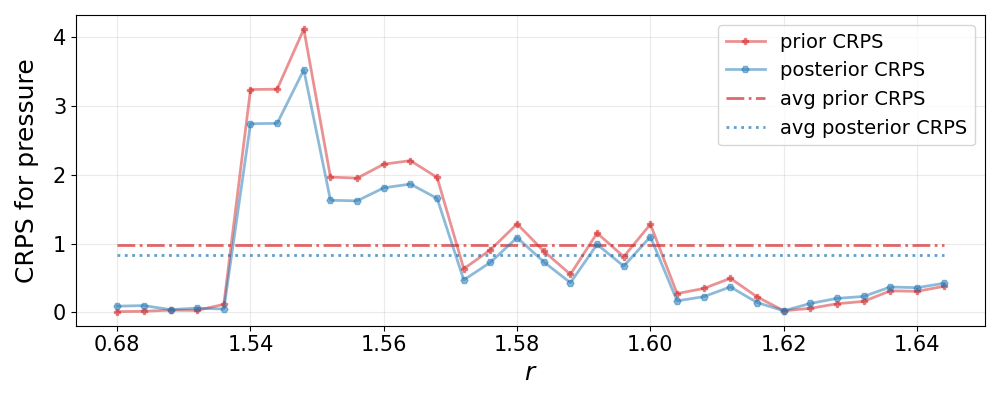}
\caption{Continuous rank probability score (CRPS) for posterior predictive versus the
  prior predictive for heat flux (top) and pressure (bottom). The dotted horizontal lines
  represent the average CRPS scores at the different observation points, $r^{\text{Exp}}_i$'s. The average CRPS scores are simply the uniform average over the different observed data points. }
\label{fig: CRPS errors}
\end{figure}

\begin{figure}[h!]
\centering
\includegraphics[width=12cm]{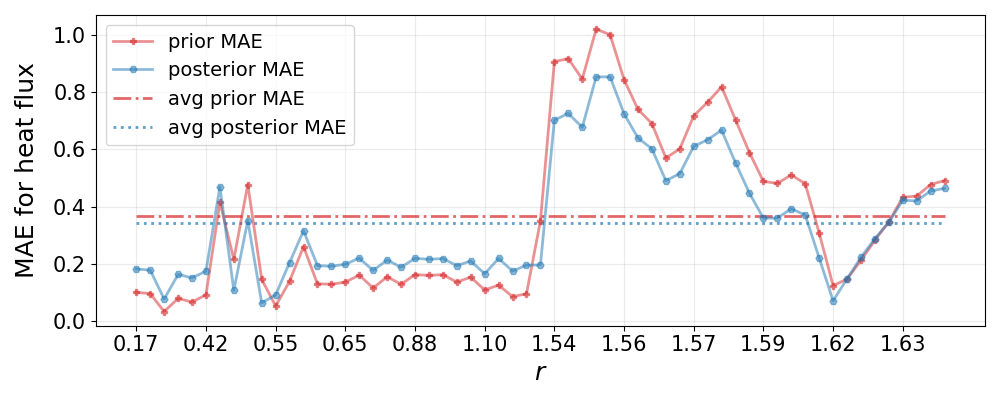}
\includegraphics[width=12cm]{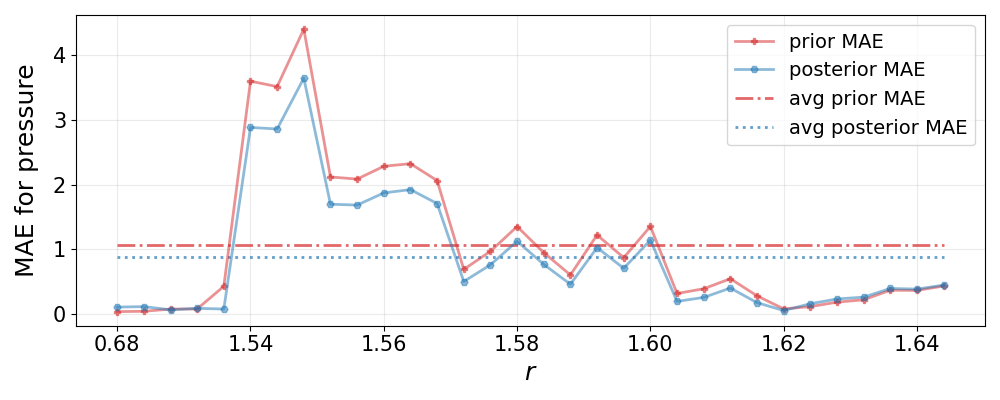}
\caption{Mean absolute errors (MAE) for posterior predictive versus the prior predictive
  for heat flux (top) and pressure (bottom). The dotted horizontal lines represent the
  average MAE scores.}
\label{fig: MAE errors}
\end{figure}

The improvement in predictive skill, post calibration, can be quantified using both the
mean absolute error (MAE;~\cite{07gb3a,07gr2a}) and continuous rank probability
score (CRPS;~\cite{07gb3a,07gr2a}). These can be computed for the predictions using
samples picked from the posterior JPDF (as illustrated in \cref{fig: posterior pred vs exp
  heat flux} and \cref{fig: posterior pred vs exp pressure}) and compared to their
counterparts computed using samples picked from the prior density \cite{Zamo2018}. Plots
of the actual distributions (not summaries) of heat flux and pressure predictions are in
Appendix~\ref{sec:bcr} (\cref{fig: prior vs posterior hist heat flux} and \cref{fig: prior
  vs posterior hist pressure}). For each of the experimental data points, we compute the
CRPS scores in \cref{fig: CRPS errors}.  Overall, the CRPS averaged over all experimental observations is reduced by the Bayesian calibration procedure compared with the prior
predictive, which is an indication of success (CRPS error is roughly the same for the the heat flux, but reduced by roughly 10\% for pressure). Likewise, if we look at the MAE averaged
over all experimental points in \cref{fig: MAE errors}, we see the same result, to a
slightly higher degree (a decrease in error of about 5\% for the heat flux and 20\% for pressure). We note that in order to average different MAE and CRPS score over
different experimental points over different scales, we weigh each of the experimental
data points by the inverse mean squared error of the magnitude of observations. The net
effect of this is that the errors can be interpreted as relative errors. 
In both cases, the calibration results seem to favor
improvement of the pressure field, as opposed to the heat flux which sees almost no change
between prior and posterior predictive results for heat flux. This is because the model
discrepancy error in the pressure field dominates the heat flux errors (see \cref{fig: LHS study} which shows how the LHS runs envelope heat flux better than pressure).  A summary of the
average CRPS and MAE errors for the prior predictive versus the posterior predictive are
shown in \cref{Tab:CRPS table} and \cref{Tab:MAPE table}.

\begin{table}[]
\begin{tabular}{|l|c|c|}
\hline
\cellcolor[HTML]{FFFFFF}{\color[HTML]{333333} \textbf{field/ prediction}} & \multicolumn{1}{l|}{\textbf{prior predictive CRPS error}} & \multicolumn{1}{l|}{\textbf{poste{}rior predictive CRPS error}} \\ \hline
\textbf{heat flux}                                                        & 0.302                                                     & 0.307                                                         \\ \hline
\textbf{pressure}                                                         & 0.976                                                     & 0.837                                                         \\ \hline
\end{tabular}
\caption{Average CRPS errors for heat flux and pressure fields for the prior predictive
  and posterior predictive densities. A decrease in CRPS is preferred.}
\label{Tab:CRPS table}
\end{table}

\begin{table}[]
\begin{tabular}{|l|c|c|}
\hline
\cellcolor[HTML]{FFFFFF}{\color[HTML]{333333} \textbf{field/ prediction}} & \multicolumn{1}{l|}{\textbf{prior predictive MAE error}} & \multicolumn{1}{l|}{\textbf{posterior predictive MAE error}} \\ \hline
\textbf{heat flux}                                                        & 0.366                                                    & 0.342                                                        \\ \hline
\textbf{pressure}                                                         & 1.072                                                    & 0.878                                                        \\ \hline
\end{tabular}
\caption{Average MAE errors for heat flux and pressure fields for the prior predictive and
  posterior predictive densities. A decrease in MAE is preferred.}
\label{Tab:MAPE table}
\end{table}

\subsection{Discussion}

The marginal posterior PDFs in \cref{fig: posterior univariate histograms} (solid lines)
show that only a handful of SST parameters can be estimated well from the heat flux and
pressure measurements. Since the heat flux and pressure measurements depend strongly on
freestream quantities, there is always a doubt whether the difficulty in estimating SST
parameters could be due to the uncertainties in the freestream quantities. Therefore, we
reran the Bayesian procedure while the freestream quantities were held at their nominal values
i.e., $\rho_s = v_s = u_s = 1$. The marginalized posterior PDFs so obtained are plotted in
\cref{fig: posterior univariate histograms} using a green dashed line. There is not a big 
change in the posterior PDFs, indicating that the inclusion of the freestream quantities
did not negatively impact the estimation problem. This is probably due to the very narrow
priors $\mathcal{U}(0.98, 1.02)$ used for $\rho_s$ and $v_s$. Temperature, as expected,
had no effect on the estimation of SST parameters.

The sharpness of the PDFs for the SST parameters could perhaps be improved by removing
some of the hard-to-estimate SST parameters (i.e., the ones whose posterior and prior PDFs
do not differ significantly). Such an exercise is left for future work, but it would,
undoubtedly, require the use of surrogate models (Sec.~\ref{sec:surrogate}) to perform
global sensitivity analysis (GSA) to choose the influential subset of SST
parameters. \cref{fig: posterior corner plot} in the Appendix~\ref{sec:bcr} shows, via
pair plots, that the SST parameters are not very correlated in the 12-dimensional JPDF,
indicating that the removal of less influential SST parameters will not adversely affect
the performance of the calibrated SST model.\footnote{The ability to easily perform sensitivity analysis using Sobol indices is another motivation for using PCEs, from which Sobol indices can be easily extracted due to their orthogonal basis representation. }

\cref{fig: posterior pred vs exp heat flux} and \cref{fig: posterior pred vs exp pressure}
show the effect of model-form errors in the SST model which prevent it from modeling the
separation zone accurately. The net effect of calibration is to improve the prediction of
pressure and heat flux in the separation zone while degrading it elsewhere. The large
number of measurements in the separation zone also contributed to the outsized importance
of this zone during calibration. It may be possible to obtain an arguably better SST model
by removing all measurements from the separation zone i.e. $r \gtrapprox 1.5$. While such
a model would not be very predictive in the separation zone, it would be highly accurate
over the cone and the cylindrical sections $0.4 \lessapprox r \lessapprox 1.5$ which
accounts for a large fraction of the heating of the HIFiRE-1 geometry.

Note that in this study we have not used the enthalpy of the incoming flow and its static
pressure, both of which were measured in the HIFiRE-1 experiment. This is because they
only help with estimation of the freestream quantities and carry no information at all
about the turbulence model.

The preceding paragraphs reveal some very useful and practical information about
hypersonic turbulence and wind-tunnel experiments. The freestream uncertainties in the
HIFiRE-1 experiment were inconsequential to the turbulence model calibration. The
information content in the heat flux and pressure measurements on the HIFiRE-1 geometry is
limited and informs only a few turbulence model parameters. However, it is the RANS's
model-form uncertainties that are the calibration bottleneck. These findings do not exist
in fluid dynamics literature for realistic hypersonic flows in engineering geometries, and
were made possible only by our ability to construct surrogate models of fields encountered
in hypersonic turbulent flows and use them within a Bayesian inference framework. The
closest equivalent to our work is Ref.~\cite{18zf2a} which targets a Mach 6.1 hypersonic turbulent flow
over a flat plate. 

\section{Summary and conclusions}
\label{sec:concl}

In this paper we formulated a surrogate which combines projection-based
model reduction techniques with machine learning regressors for the prediction of
scalar-valued fields. We used principal component analysis to perform the dimension
reduction and then explored a variety of different machine learning regressors to fit the
projection coefficients of the learned components. We experimented with Gaussian process
regression, polynomial chaos expansions, random forests, kernel ridge regression, support
vector machines, and multi-layer perceptron models. In order to tune each scalar regressor
over a given set of hyperparameters and perform model selection, k-fold cross validation
was used. Ultimately, for the final surrogate, a multivariate polynomial representation,
i.e., a polynomial chaos expansion with multivariate Legendre polynomials, was chosen to
fit our reduced space projection coefficients. The ML experiments showed that polynomials
provided the greatest amount of expressivity and accuracy, while being
the easiest and simplest to train. They also provided the most consistent answers (e.g., without dependence on say random seeds like MLPs) which is critical for  reproducibility. We demonstrated the efficacy and accuracy of
these surrogates for predicting the heat flux and pressure fields on the surface of the HIFiRE
geometry in a Mach 7.16 turbulent flow using $m=2500$ simulation runs. The surrogate was then used in a first-ever Bayesian calibration of the HIFiRE
experiment, using an adaptive MCMC method to construct a joint density. The posterior predictive
samples from the JPDF matched the experiment data better than the prior predictive samples
for both heat flux and the pressure fields in terms of both the CRPS (continuous rank
probability score) and MAE (mean absolute error), thus resulting in an improved predictive model and reduced mismatch between prediction and experimental data.

The Bayesian calibration of the SST model parameters was also 
able to construct posterior PDFs, compare them with the prior and discern which SST model
parameters could be estimated well from the heat flux and pressure measurements. We
discovered that the range over which the freestream quantities were controlled in the
LENS-I wind-tunnel during the HIFiRE-1 experiment was sufficiently narrow that the
uncertainty did not impact the turbulence model estimation problem. The limiting factor
was the model-form error in the RANS model which made it infeasible to capture the
separation zone at the extreme aft of the test geometry. Further improvement
in the calibrated model may be had by performing a GSA to pick the most sensitive SST
model parameters and calibrating them to the same dataset. These findings are novel and
were made possible by the numerical and statistical tools developed in this paper. In
addition, the same tools can be used to perform the GSA.

Last, but not least, we have provided software, \emph{tesuract} (Tensor Surrogate Automation and Computation), to build the types of surrogates used in this paper. \emph{tesuract} is
built on top of the scikit-learn API \cite{scikit-learn} and utilizes scikit-learn's vast
library of machine learning estimators, model selection techniques, and dimension
reduction methods in a unique ML pipeline which allows the construction of surrogates for
both single target scalar outputs and scalar-valued fields (i.e., multi-target correlated
outputs). This allows flexibility, utility and easy-of-use for many applications related
to surrogate construction. 
This software is freely available on github (\hyperlink{https://github.com/kennychowdhary/tesuract}{https://github.com/kennychowdhary/tesuract}).


\section{Funding and Acknowledgment}

Sandia National Laboratories is a multimission laboratory managed and operated by National
Technology and Engineering Solutions of Sandia, LLC., a wholly owned subsidiary of
Honeywell International, Inc., for the U.S. Department of Energy's National Nuclear
Security Administration under contract DE-NA0003525. his paper describes objective
technical results and analysis. Any subjective views or opinions that might be expressed
in the paper do not necessarily represent the views of the U.S. Department of Energy or
the United States Government.

\appendix
\section{Final ROM machine learning models}
\label{appendix: rom models}

Here we discuss the hyper-parameter tuning procedure and the final ROM-based
model architectures for both heat flux and pressure fields. Recall that we
have chosen four PCA components for our reduced order model. The
corresponding projection coefficients associated for each of the four
components is then fit with six different machine learning regressor models, and
each of these regressors is hyper-parameter tuned over a set of possible
parameter combinations. These parameter combinations are listed below, of which 
more detail can be found in the documentation of our
surrogate construction software \textit{tesuract} and sklearn's website. A
5-fold cross validation score was computed for every single combination of
grid values.  
\begin{verbatim}
# polynomial chaos regressor
pce_grid = {
  'order': list(range(1,11)),
  'mindex_type': ['total_order','hyperbolic'],
  'fit_type': ['LassoCV','ElasticNetCV','linear']}

# random forest regressor
rf_grid = {
  'n_estimators': [200,500,1000,5000],
  'max_features': [3,'sqrt','auto'],
  'max_depth': [5,10,50]}

# multi-layer perceptron regressor
mlp_grid = {
  'hidden_layer_sizes': [(50,),(50,)*2,(50,)*4,(50,)*6
                         (100,),(100,)*2,(100,)*4,(100,)*6
                         (500,),(500,)*2,(500,)*4,(500,)*6],
  'solver': ['lbfgs','adam','sgd'],
  'activation': ['relu'],
  'max_iter': [2500],
  'batch_size': ['auto'],
  'learning_rate': ['invscaling'],
  'alpha': [1e-4,1e-6,1e-2],
  'tol': [1e-6,1e-4],
  'random_state': [0,99,324]}

# kernel ridge regression regressor
krr_grid = {
  'kernel': ['polynomial'],
  'kernel_params': [{'degree':1},{'degree':2},{'degree':3},{'degree':4}],
  'alpha': [1e-4,1e-2,1e-1,1.0]}

# gaussian process regressor
gpr_grid = {
  'kernel': [1.0 * RBF(.1) + .1**2 *WhiteKernel(.1),
             1.0 * RBF(.1) + .1**2 *WhiteKernel(.1) + 1.0*DotProduct(.1), 
             1.0 * Matern(length_scale=.1, nu=1.5) + .1**2 *WhiteKernel(.1)],
  'alpha': [1e-10],
  'optimizer': ['fmin_l_bfgs_b'],
  'n_restarts_optimizer': [2],
  'random_state': [0]}

# k-nearest neighbor regressor
knn_grid = {
  'n_neighbors': (1,5,8,10),
  'leaf_size': (20,30,40,1),
  'p': (1,2),
  'weights': ('uniform', 'distance'),
  'metric': ('minkowski', 'chebyshev')}     

# support vector machine regressor
svr_grid = {
  'kernel': ('linear','poly', 'rbf', 'sigmoid'),
  'degree': (2,4,8),
  'gamma': ('scale','auto'),
  'C': (1,5,10)} 
\end{verbatim}
To give an example, consider the the polynomial chaos regressor parameter grid
above. A 5-fold cross validation score was computed for each and every combination of
the \verb|order|, the polynomial total degree, \verb|mindx_type|, which controls the number of interaction terms, and \verb|fit_type|, the algorithm for solving the least squares problem. There are 60 parameter combinations and, thus, a total of 300 PCE fits were computed (five for each of the 60 parameter combinations since we are using five-fold cross-validation) for each of the four components. This method was repeated for each of the regressors listed above in order to obtain the penultimate model for both heat flux and pressure fields. Thus, each model comparison involves hundreds or thousands of ML regression fits, all of which is handled neatly and efficiently within the tesuract and the sklearn framework so that the user does not need to bother with the cumbersome nesting and splitting of the test and train data. 

The PCE and MLP models had the highest cross-validation scores among the
regressors. For heat flux, the PCE model with the highest cross-validation
score had polynomial orders of degrees $\{2,2,4,4\}$ for each of the four
components, and for pressure the PCE model with the highest cross-validation
score had polynomial orders of $\{4,4,4,4\}$. In contrast, the best MLP
network for each projection coefficient had 4 hidden layers of 50 nodes each
for both the pressure and heat flux fields. The network used rectified linear
units for the activation functions, a tolerance of $10^{-6}$ for the LBFGS
solver, and an inverse scaling for the learning rate, which gradually
decreases as the time step progresses. The rest of the parameters were set to
their default values \cite{scikit-learn}. 

\section{Bayesian calibration results}
\label{sec:bcr}

\begin{figure}[h!]
\centering
\includegraphics[width=8cm]{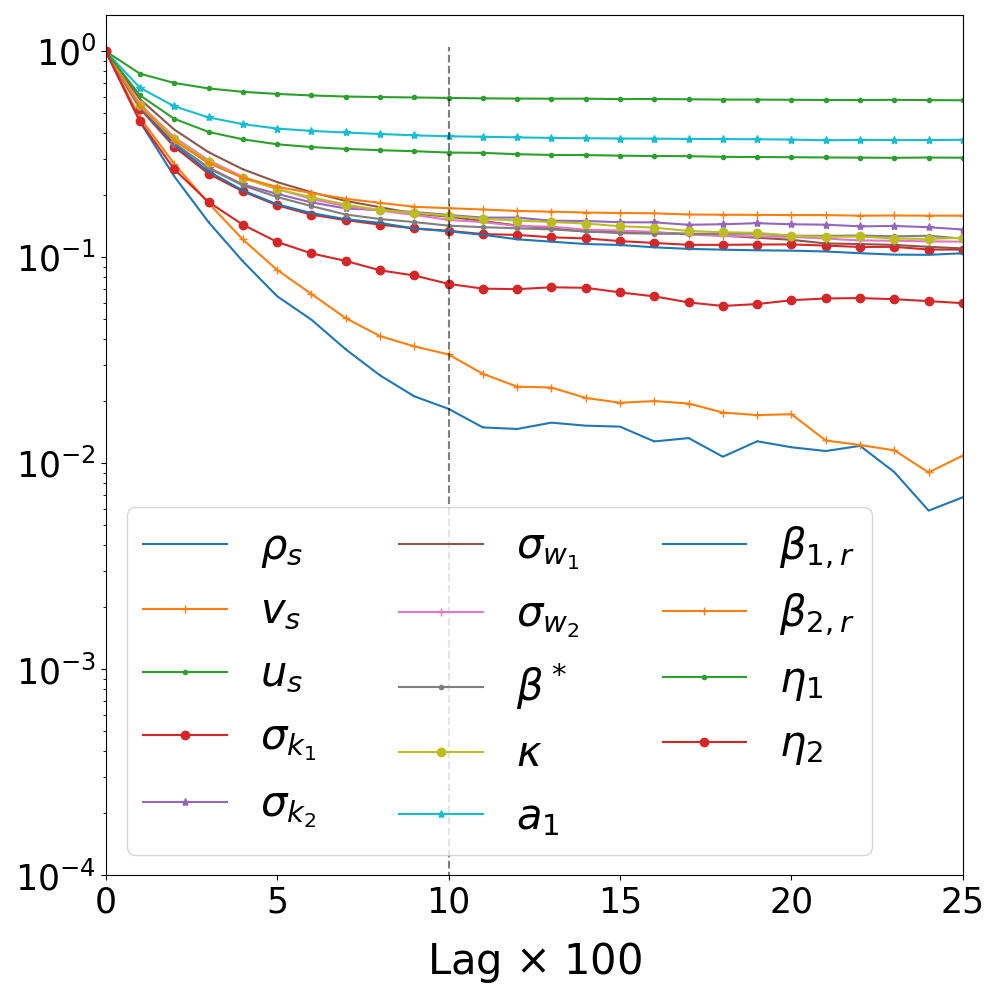}
\caption{Autocorrelation time for each of the twelve calibration parameters, including the
  two noise parameters, $\eta_1,\eta_2$. The horizontal dotted line represents the amount
  by which we thin the chain, giving us an effective sample size of roughly 24k samples in
  our twelve-dimensional sample space (plus two noise parameters, i.e., $\Gamma$
  distributed random variables $\eta_1$ and $\eta_2$ representing the inverse variance of
  the model discrepancy errors).}
\label{fig: autocor lag time}
\end{figure}

The adaptive MCMC algorithm described in Sec.~\ref{sec:calib} yielded a chain of that was thinned to
reduce the autocorrelation in the sequence of samples. The autocorrelation vs lag time is plotted in
\cref{fig: autocor lag time}. \footnote{The aggressive thinning accounted for the
  variability in the autocorrelation amongst the different parameters and was chosen so
  that all twelve parameters has sufficiently converged according to the autocorrelation
  lag plot in \cref{fig: autocor lag time}.}

\begin{figure}[h!]
\centering
\includegraphics[width=1.0\textwidth]{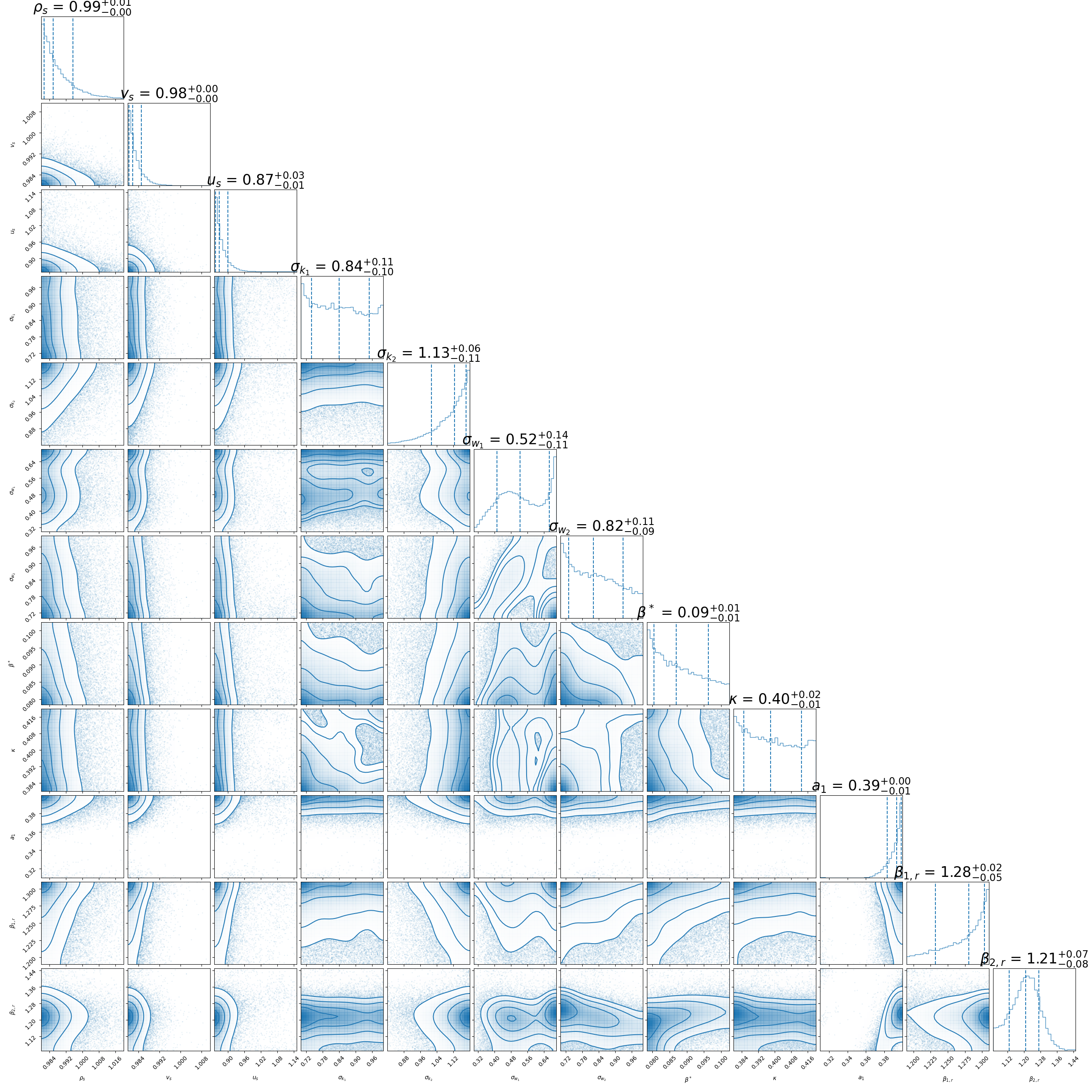}
\caption{Corner (pair) plot for the posterior distribution over the twelve calibration or
  tuning parameters. The maximum \emph{a posteriori} estimate is displayed above the
  univariate plots. Given that the priors are chosen to be uniform, all parameters are
  informed, to some extent,  from the observed data. }
\label{fig: posterior corner plot}
\end{figure}

The full JPDF, illustrated as a matrix of pair plots, is shown in
\cref{fig: posterior corner plot}. While \cref{fig: posterior univariate histograms} shows
the SST parameters that could be estimated from the HIFiRE-1 measurements, \cref{fig:
  posterior corner plot} shows the correlations that exist between the various SST
parameters in the posterior JPDF. We see that the correlations are mild i.e., the
structures in the 2D plots are mostly horizontally or vertically aligned. This is
fortunate as it implies that the SST parameters that cannot be inferred well (i.e., where
prior and posterior PDFs in \cref{fig: posterior univariate histograms} are similar) can
be simply removed to yield a smaller estimation problem without materially (negatively)
impacting the accuracy of the SST turbulence model.

\begin{figure}[h!]
\centering
\includegraphics[width=0.8\textwidth]{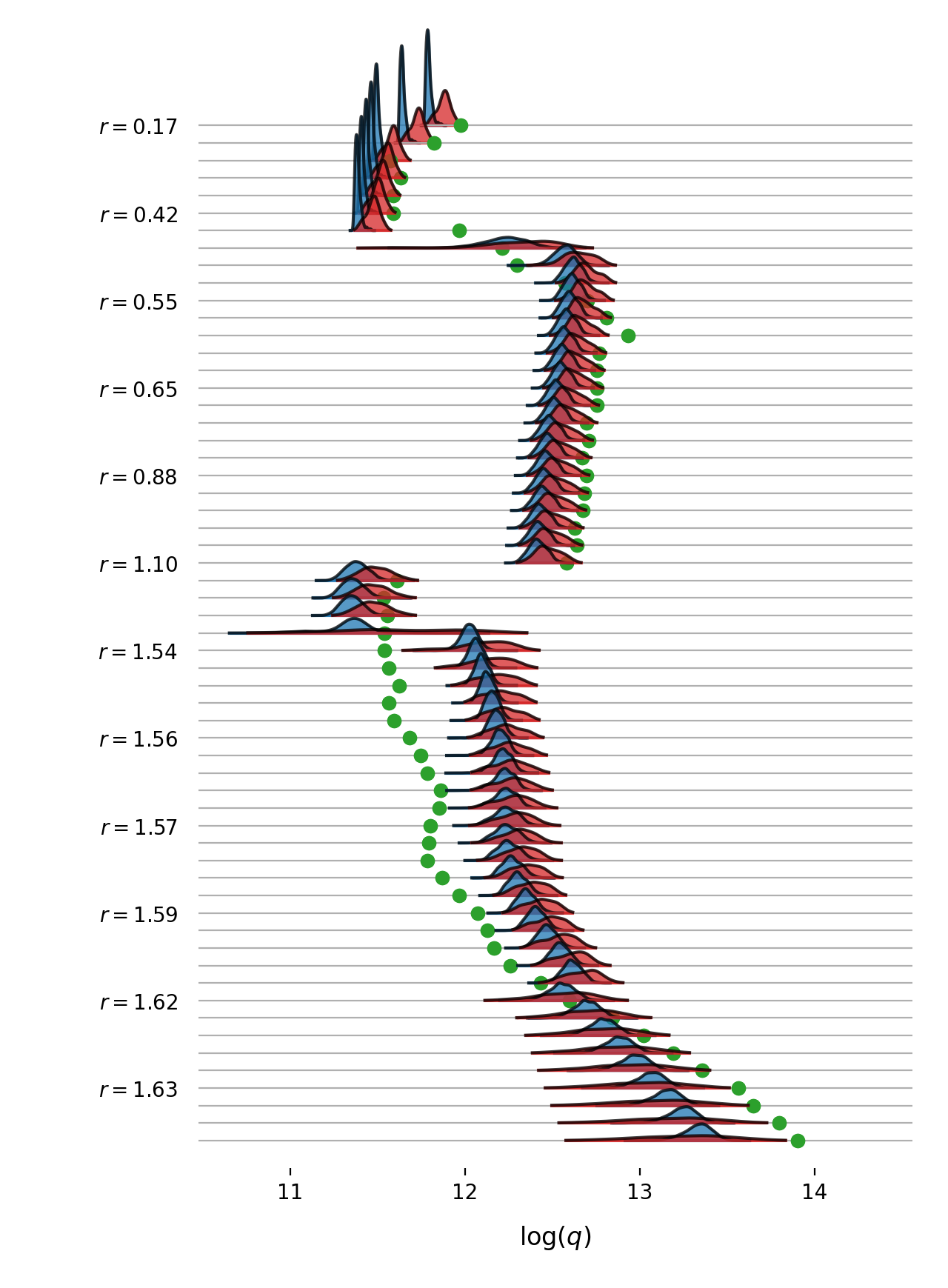}
\caption{Prior versus the posterior predictive distributions for log-scaled heat flux,
  denoted by $\log (q)$. $r$-axis is displayed on the vertical, the blue histogram
  represents the posterior predictive, while the red shows the prior. The green dot shows
  the single observed data.}
\label{fig: prior vs posterior hist heat flux}
\end{figure}

The posterior predictive densities, computed by simulating the HIFiRE-1 experiments with
${\mathbf{x}}$ drawn from the posterior JPDF (as plotted in \cref{fig: posterior corner
  plot} and marginalized in \cref{fig: posterior univariate histograms}) are plotted in
\cref{fig: prior vs posterior hist heat flux} (heat flux) and \cref{fig: prior vs
  posterior hist pressure} (pressure) in blue. The prior predictive are plotted in
red. These predictive densities were used to compute the CRPS and MAE in \cref{fig: CRPS
  errors} and \cref{fig: MAE errors}. These plots are colloquially known as \textit{joy
  plots}. 
The y-axis represents the locations of the observed data point and the x-axis represents
the $\log$ values of the heat flux and/or pressure.
  
\begin{figure}[h!]
\centering
\includegraphics[width=0.8\textwidth]{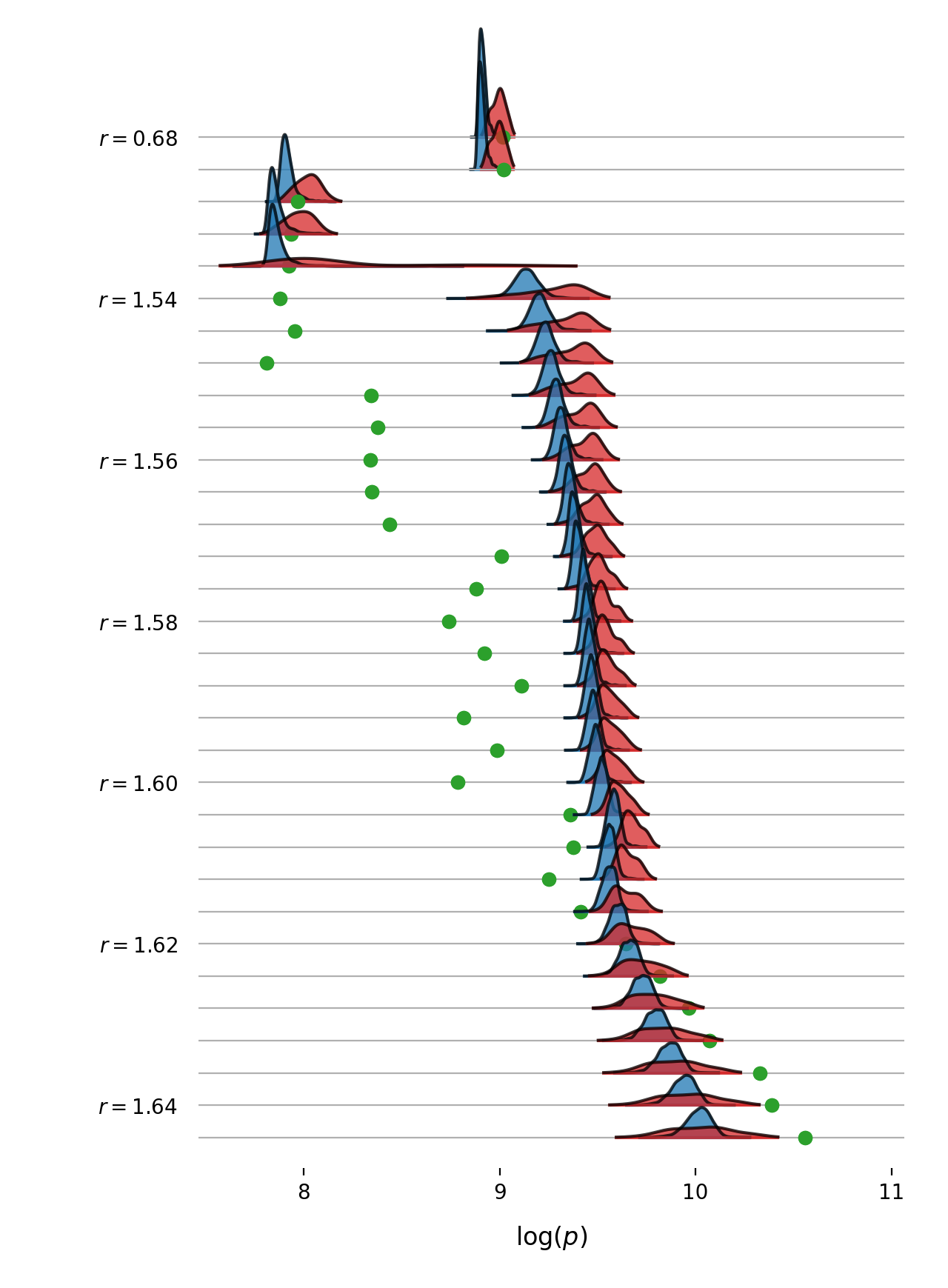}
\caption{Prior versus the posterior predictive distributions for log-scaled pressure,
  denoted by $\log (p)$. $r$-axis is displayed on the vertical, the blue histogram
  represents the posterior predictive, while the red shows the prior. The green dot shows
  the single observed data.}
\label{fig: prior vs posterior hist pressure}
\end{figure}

\clearpage
\bibliographystyle{aiaa}
\bibliography{paper}

\end{document}

%% file: front_matter.tex
\usepackage[table,xcdraw]{xcolor}
\usepackage[english]{babel}
\usepackage{amssymb,bbm,amsmath,graphicx}

\usepackage{algorithm}
\usepackage{algorithmic}

\usepackage{amsmath}

\usepackage{amsfonts}

\usepackage{bm}

\usepackage{braket}

\usepackage{tabularx}

\usepackage{pgfplots}
\pgfplotsset{compat=1.12}

\usepackage{lineno}

\usepackage[margin=1.25in]{geometry}
\usepackage[font=footnotesize,justification=Centering,singlelinecheck=true]{subfig}

\usepackage{xcolor}

\DeclareMathOperator*{\argmin}{arg\,min}

\usepackage{hyperref}
\hypersetup{
  colorlinks=true,
  allcolors=green!50!black,
  urlcolor=red!50!black
}

\usepackage[capitalize,nameinlink]{cleveref}

\usepackage{bm}

\usepackage{comment}
\usepackage[normalem]{ulem}	

%% file: incl.tex
\newcommand{\Minf}{M_{\infty}}
\newcommand{\rbDim}{n}




%% file: paper.bbl
\begin{thebibliography}{10}
\newcommand{\enquote}[1]{``#1''}

\bibitem{20aa3a}
Alizadeh, R., Allen, J.~K., and Mistree, F., \enquote{Managing computational
  complexity using surrogate models: a critical review,} {\em Research in
  Engineering Design\/}, Vol.~31, 2020, pp.~275--298.

\bibitem{18ya3a}
Yondo, R., Andrés, E., and Valero, E., \enquote{A review on design of
  experiments and surrogate models in aircraft real-time and many-query
  aerodynamic analyses,} {\em Progress in Aerospace Sciences\/}, Vol.~96, 2018,
  pp.~23--61.

\bibitem{17sm3a}
Sudret, B., Marelli, S., and Wiart, J., \enquote{Surrogate models for
  uncertainty quantification: An overview,} {\em 2017 11th European Conference
  on Antennas and Propagation (EUCAP)\/}, 2017, pp. 793--797.

\bibitem{15ac4a}
Asher, M.~J., Croke, B.~F.~W., Jakeman, A.~J., and Peeters, L.~J.~M.,
  \enquote{A review of surrogate models and their application to groundwater
  modeling,} {\em Water Resources Research\/}, Vol.~51, No.~8, 2015,
  pp.~5957--5973.

\bibitem{21kz2a}
Kou, J. and Zhang, W., \enquote{Data-driven modeling for unsteady aerodynamics
  and aeroelasticity,} {\em Progress in Aerospace Sciences\/}, Vol.~125, 2021,
  pp.~100725.

\bibitem{19sm4a}
Swischuk, R., Mainini, L., Peherstorfer, B., and Willcox, K.,
  \enquote{Projection-based model reduction: Formulations for physics-based
  machine learning,} {\em Computers and Fluids\/}, Vol.~179, 2019,
  pp.~704--717.

\bibitem{hoang2021projection}
Hoang, C., Chowdhary, K., Lee, K., and Ray, J., \enquote{Projection-based model
  reduction of dynamical systems using space-time subspace and machine
  learning,} {\em arXiv preprint arXiv:2102.03505\/}, 2021.

\bibitem{08wm4a}
Wadhams, T.~P., Mundy, E., MacLean, M.~G., and Holden, M.~S., \enquote{Ground
  test studies of the HIFiRE-1 transition experiment Part 1: Experimental
  results,} {\em Journal of Spacecraft and Rockets\/}, Vol.~45, No.~6, 2008,
  pp.~1134--1148.

\bibitem{08mw4a}
MacLean, M., Wadhams, T., Holden, M., and Johnson, H., \enquote{Ground test
  studies of the HIFiRE-1 transition experiment Part 2: Computational
  analysis,} {\em Journal of Spacecraft and Rockets\/}, Vol.~45, No.~6, 2008,
  pp.~1149--1164.

\bibitem{06wd1a}
Wilcox, D.~C., {\em Turbulence Modeling for CFD\/}, D C W Industries, La
  Canada, CA, USA, 3rd ed., 2006.

\bibitem{94mf1a}
Menter, F.~R., \enquote{Two-equation eddy-viscosity turbulence models for
  engineering applications,} {\em AIAA Journal\/}, Vol.~32, No.~8, August 1994,
  pp.~1598--1605.

\bibitem{14gg3a}
Guillas, S., Glover, N., and Malki-Epshtein, L., \enquote{Bayesian calibration
  of the constants of the k-$\epsilon$ turbulence model for a CFD model of
  street canyon flow,} {\em Computer Methods in Applied Mechanics and
  Engineering\/}, Vol.~279, 2014, pp.~536--553.

\bibitem{18rd5a}
Ray, J., Dechant, L., Lefantzi, S., Ling, J., and Arunajatesan, S.,
  \enquote{Robust Bayesian calibration of a k-$\epsilon$ model for compressible
  jet-in-crossflow simulations,} {\em AIAA Journal\/}, Vol.~56, No.~12, 2018,
  pp.~4893--4909.

\bibitem{19zf2a}
Zhang, J. and Fu, S., \enquote{An efficient approach for quantifying parameter
  uncertainty in the SST turbulence model,} {\em Computers and Fluids\/},
  Vol.~181, 2019, pp.~173--187.

\bibitem{21vs2a}
Viana, F.~A.~C. and Subramaniyan, A.~K., \enquote{A Survey of Bayesian
  Calibration and Physics-informed Neural Networks in Scientific Modeling,}
  {\em Archives of Computational Methods in Engineering\/}, Vol.~28, 2021,
  pp.~3801–3830.

\bibitem{08my2a}
McPhee, J. and Yeh, W. W.-G., \enquote{Groundwater Management Using Model
  Reduction via Empirical Orthogonal Functions,} {\em Journal of Water
  Resources Planning and Management\/}, Vol.~134, No.~2, 2008, pp.~161--170.

\bibitem{89dw2a}
Dunbar, W.~S. and Woodbury, A.~D., \enquote{Application of the Lanczos
  Algorithm to the solution of the groundwater flow equation,} {\em Water
  Resources Research\/}, Vol.~25, No.~3, 1989, pp.~551--558.

\bibitem{90wd3a}
Woodbury, A.~D., Dunbar, W.~S., and Nour-Omid, B., \enquote{Application of the
  Arnoldi Algorithm to the solution of the advection-dispersion equation,} {\em
  Water Resources Research\/}, Vol.~26, No.~10, 1990, pp.~2579--2590.

\bibitem{05wm2a}
Willcox, K. and Megretski, A., \enquote{Fourier Series for Accurate, Stable,
  Reduced-Order Models in Large-Scale Linear Applications,} {\em SIAM Journal
  on Scientific Computing\/}, Vol.~26, No.~3, 2005, pp.~944--962.

\bibitem{08gw2a}
Gugercin, S. and Willcox, K., \enquote{Krylov projection framework for Fourier
  model reduction,} {\em Automatica\/}, Vol.~44, No.~1, 2008, pp.~209--215.

\bibitem{ly2001modeling}
Ly, H.~V. and Tran, H.~T.,
  \enquote{\href{https://www.sciencedirect.com/science/article/pii/S0895717700002405}{Modeling
  and control of physical processes using proper orthogonal decomposition},}
  {\em Mathematical and computer modelling\/}, Vol.~33, No. 1-3, 2001,
  pp.~223--236.

\bibitem{higdon2008computer}
Higdon, D., Gattiker, J., Williams, B., and Rightley, M.,
  \enquote{\href{https://www.tandfonline.com/doi/abs/10.1198/016214507000000888}{Computer
  model calibration using high-dimensional output},} {\em Journal of the
  American Statistical Association\/}, Vol.~103, No. 482, 2008, pp.~570--583.

\bibitem{audouze2009reduced}
Audouze, C., De~Vuyst, F., and Nair, P.,
  \enquote{\href{https://onlinelibrary.wiley.com/doi/abs/10.1002/nme.2540?casa_token=utLvPmeWZpYAAAAA:TvAmGXkwD8qasAD3t4LPLMczOioxEUyrmJTAl8dPyxaImjraSaEN10QbpUaZicaVcEiqQvRVN_b5cPhz}{Reduced-order
  modeling of parameterized PDEs using time--space-parameter principal
  component analysis},} {\em International journal for numerical methods in
  engineering\/}, Vol.~80, No.~8, 2009, pp.~1025--1057.

\bibitem{audouze2013nonintrusive}
Audouze, C., De~Vuyst, F., and Nair, P.~B.,
  \enquote{\href{https://onlinelibrary.wiley.com/doi/full/10.1002/num.21768?casa_token=CTeWEmFKDB8AAAAA\%3AbnTdsc6R8YyCWWS5RjX9F4W3Ujaae6xZAMCKO-_nbVWLGNWzjKvSx7BWWRH0u30dxARyhwdKgBVdhpxD}{Nonintrusive
  reduced-order modeling of parametrized time-dependent partial differential
  equations},} {\em Numerical Methods for Partial Differential Equations\/},
  Vol.~29, No.~5, 2013, pp.~1587--1628.

\bibitem{wirtz2015surrogate}
Wirtz, D., Karajan, N., and Haasdonk, B.,
  \enquote{\href{https://onlinelibrary.wiley.com/doi/full/10.1002/nme.4767?casa_token=jLXRM09Hm_0AAAAA\%3AyZeLiasySg5KB1JrFr9J-4vZS75AFIlv84jIKWQwrB-lv0211qO0uMp4MH_wfAXivwHVqufFDrReSe9L}{Surrogate
  modeling of multiscale models using kernel methods},} {\em International
  Journal for Numerical Methods in Engineering\/}, Vol.~101, No.~1, 2015,
  pp.~1--28.

\bibitem{mainini2015surrogate}
Mainini, L. and Willcox, K.,
  \enquote{\href{https://arc.aiaa.org/doi/full/10.2514/1.J053464?casa_token=4rfVZAXT1xsAAAAA\%3Ap2JLNGcU8cDE0EGzzdTX_jVIbV1HuZ6HjL9hsL_ShaQEG9prLeMtIc6ImfIlryNfZiAYeYmKddQ}{Surrogate
  modeling approach to support real-time structural assessment and decision
  making},} {\em AIAA Journal\/}, Vol.~53, No.~6, 2015, pp.~1612--1626.

\bibitem{ulu2016data}
Ulu, E., Zhang, R., and Kara, L.~B.,
  \enquote{\href{https://www.tandfonline.com/doi/abs/10.1080/21681163.2015.1030775}{A
  data-driven investigation and estimation of optimal topologies under variable
  loading configurations},} {\em Computer Methods in Biomechanics and
  Biomedical Engineering: Imaging \& Visualization\/}, Vol.~4, No.~2, 2016,
  pp.~61--72.

\bibitem{hesthaven2018non}
Hesthaven, J.~S. and Ubbiali, S.,
  \enquote{\href{https://www.sciencedirect.com/science/article/pii/S0021999118301190?casa_token=nO5FSa3zTzYAAAAA:pp5fTfDBEuPTxWqmAC0nchM7uGaPK4BwqtsJdXoaWjQe2q_WTQGGEUM5UZwaYaLS_r_hm19thwk}{Non-intrusive
  reduced order modeling of nonlinear problems using neural networks},} {\em
  Journal of Computational Physics\/}, Vol.~363, 2018, pp.~55--78.

\bibitem{20mm6a}
Maulik, R., Mohan, A., Lusch, B., Madireddy, S., Balaprakash, P., and Livescu,
  D., \enquote{Time-series learning of latent-space dynamics for reduced-order
  model closure,} {\em Physica D: Nonlinear Phenomena\/}, Vol.~405, Apr 2020,
  pp.~132368.

\bibitem{19rp5a}
Rahman, S.~M., Pawar, S., San, O., Rasheed, A., and Iliescu, T.,
  \enquote{Nonintrusive reduced order modeling framework for quasigeostrophic
  turbulence,} {\em Phys. Rev. E\/}, Vol.~100, Nov 2019, pp.~053306.

\bibitem{18wx6a}
Wang, Z., Xiao, D., Fang, F., Govindan, R., Pain, C.~C., and Guo, Y.,
  \enquote{Model identification of reduced order fluid dynamics systems using
  deep learning,} {\em International Journal for Numerical Methods in
  Fluids\/}, Vol.~86, 2018, pp.~255--268.

\bibitem{19mz5a}
Mo, S., Zhu, Y., Zabaras, N., Shi, X., and Wu, J., \enquote{Deep Convolutional
  Encoder-Decoder Networks for Uncertainty Quantification of Dynamic Multiphase
  Flow in Heterogeneous Media,} {\em Water Resources Research\/}, Vol.~55,
  No.~1, 2019, pp.~703--728.

\bibitem{21lp2a}
Lee, K. and Parish, E.~J., \enquote{Parameterized neural ordinary differential
  equations: applications to computational physics problems,} {\em Proceedings
  of the Royal Society A: Mathematical, Physical and Engineering Sciences\/},
  Vol.~477, No. 2253, 2021, pp.~20210162.

\bibitem{chen2018greedy}
Chen, W., Hesthaven, J.~S., Junqiang, B., Qiu, Y., Yang, Z., and Tihao, Y.,
  \enquote{\href{https://arc.aiaa.org/doi/full/10.2514/1.J056161?casa_token=kRQpoBgK6rwAAAAA\%3AZTYrKpThZkF3hLENSWZ54VW-TYTB2MAMG4pmM4BILlM52oh-s7FnZY-6c970N3uMCEkEX9fVq5M}{Greedy
  nonintrusive reduced order model for fluid dynamics},} {\em AIAA Journal\/},
  Vol.~56, No.~12, 2018, pp.~4927--4943.

\bibitem{17fs5a}
Fang, J., Sun, G., Qiu, N., Kim, N.~H., and Li, Q., \enquote{On design
  optimization for structural crashworthiness and its state of the art,} {\em
  Structural and Multidisciplnary Optimization\/}, Vol.~55, 2017,
  pp.~1091–1119.

\bibitem{17dm3a}
Dey, S., Mukhopadhyay, T., and Adhikari, S., \enquote{Metamodel based
  high-fidelity stochastic analysis of composite laminates: A concise review
  with critical comparative assessment,} {\em Composite Structures\/},
  Vol.~171, 2017, pp.~227--250.

\bibitem{19ll4a}
Laurent, L., Riche, R.~L., Soulier, B., and Boucard, P., \enquote{An Overview
  of Gradient-Enhanced Metamodels with Applications,} {\em Archive of
  Computational Methods in Engineering\/}, Vol.~26, 2019, pp.~61–106.

\bibitem{swischuk2019projection}
Swischuk, R., Mainini, L., Peherstorfer, B., and Willcox, K.,
  \enquote{\href{https://www.sciencedirect.com/science/article/pii/S0045793018304250}{Projection-based
  model reduction: Formulations for physics-based machine learning},} {\em
  Computers \& Fluids\/}, Vol.~179, 2019, pp.~704--717.

\bibitem{19cn5a}
Cao, C., Nie, C., Pan, S., Cai, J., and Qu, K., \enquote{A constrained
  reduced-order method for fast prediction of steady hypersonic flows,} {\em
  Aerospace Science and Technology\/}, Vol.~91, 2019, pp.~679--690.

\bibitem{21dg4a}
Dreyer, E.~R., Grier, B.~J., McNamara, J.~J., and Orr, B.~C., \enquote{Rapid
  Steady-State Hypersonic Aerothermodynamic Loads Prediction Using Reduced
  Fidelity Models,} {\em Journal of Aircraft\/}, Vol.~58, No.~3, 2021,
  pp.~663--676.

\bibitem{15cl4a}
Chen, X., Liu, L., Long, T., and Yue, Z., \enquote{A reduced order
  aerothermodynamic modeling framework for hypersonic vehicles based on
  surrogate and POD,} {\em Chinese Journal of Aeronautics\/}, Vol.~28, No.~5,
  2015, pp.~1328--1342.

\bibitem{17cz4a}
Chen, X., Zuo, G., Shi, Y., and Liu, L., {\em An Efficient Integrated
  Aerothermoelasticity Analysis System Based on Surrogate-based Reduced Order
  Modeling for Hypersonic Vehicles\/}, 2017.

\bibitem{19cz2a}
Chen, Z. and Zhao, Y., \enquote{Aerothermoelastic Analysis of a Hypersonic
  Vehicle Based on Thermal Modal Reconstruction,} {\em International Journal of
  Aerospace Engineering\/}, Vol.~2019, 2019, Article ID 8384639.

\bibitem{12cm2a}
Crowell, A.~R. and McNamara, J.~J., \enquote{Model Reduction of Computational
  Aerothermodynamics for Hypersonic Aerothermoelasticity,} {\em AIAA
  Journal\/}, Vol.~50, No.~1, 2012, pp.~74--84.

\bibitem{19xj4a}
Xiaoxuan, Y., Jinglong, H., Bing, Z., and Haiwei, Y., \enquote{Model reduction
  of aerothermodynamic for hypersonic aerothermoelasticity based on POD and
  Chebyshev method,} {\em Proceedings of the Institution of Mechanical
  Engineers, Part G: Journal of Aerospace Engineering\/}, Vol.~233, No.~10,
  2019, pp.~3734--3748.

\bibitem{21zy5a}
Zhang, K., Yao, J., He, Z., Xin, J., and Fan, J., \enquote{Probabilistic
  Transient Heat Conduction Analysis Considering Uncertainties in Thermal Loads
  Using Surrogate Model,} {\em Journal of Spacecraft and Rockets\/}, Vol.~58,
  No.~4, 2021, pp.~1030--1042.

\bibitem{17vb3a}
Vollant, A., Balarac, G., and Corre, C., \enquote{Subgrid-scale scalar flux
  modelling based on optimal estimation theory and machine-learning
  procedures,} {\em Journal of Turbulence\/}, Vol.~18, No.~9, 2017,
  pp.~854--878.

\bibitem{19md2a}
Matai, R. and Durbin, P., \enquote{Large-eddy simulation of turbulent flow over
  a parametric set of bumps,} {\em Journal of Fluid Mechanics\/}, Vol.~866,
  2019, pp.~503–525.

\bibitem{19di3a}
Duraisamy, K., Iaccarino, G., and Xiao, H., \enquote{Turbulence Modeling in the
  Age of Data,} {\em Annual Review of Fluid Mechanics\/}, Vol.~51, No.~1, 2019,
  pp.~357--377.

\bibitem{19xc2a}
Xiao, H. and Cinnella, P., \enquote{Quantification of model uncertainty in RANS
  simulations: A review,} {\em Progress in Aerospace Sciences\/}, Vol.~108,
  2019, pp.~1--31.

\bibitem{19zw4a}
Zhang, X., Wu, J., Coutier-Delgosha, O., and Xiao, H., \enquote{Recent progress
  in augmenting turbulence models with physics-informed machine learning,} {\em
  Journal of Hydrodynamics\/}, Vol.~31, 2019, pp.~1153–1158.

\bibitem{16lk3a}
Ling, J., Kurzawski, A., and Templeton, J., \enquote{Reynolds-averaged
  turbulence modelling using deep neural networks with embedded invariance,}
  {\em Journal of Fluid Mechanics\/}, Vol.~807, 2016, pp.~155–166.

\bibitem{21zd6a}
Zhang, Y., Dwight, R.~P., Schmelzer, M., Gómez, J.~F., hua Han, Z., and
  Hickel, S., \enquote{Customized data-driven RANS closures for bi-fidelity
  LES–RANS optimization,} {\em Journal of Computational Physics\/}, Vol.~432,
  2021, pp.~110153.

\bibitem{19zz4a}
Zhu, L., Zhang, W., Kou, J., and Liu, Y., \enquote{Machine learning methods for
  turbulence modeling in subsonic flows around airfoils,} {\em Physics of
  Fluids\/}, Vol.~31, No.~1, 2019, pp.~015105.

\bibitem{16ws2a}
Weatheritt, J. and Sandberg, R., \enquote{A novel evolutionary algorithm
  applied to algebraic modifications of the RANS stress–strain relationship,}
  {\em Journal of Computational Physics\/}, Vol.~325, 2016, pp.~22--37.

\bibitem{20za5a}
Zhao, Y., Akolekar, H.~D., Weatheritt, J., Michelassi, V., and Sandberg, R.~D.,
  \enquote{RANS turbulence model development using CFD-driven machine
  learning,} {\em Journal of Computational Physics\/}, Vol.~411, 2020,
  pp.~109413.

\bibitem{17ws2a}
Weatheritt, J. and Sandberg, R., \enquote{The development of algebraic stress
  models using a novel evolutionary algorithm,} {\em International Journal of
  Heat and Fluid Flow\/}, Vol.~68, 2017, pp.~298--318.

\bibitem{20sd3a}
Schmelzer, M., Dwight, R.~P., and Cinnella, P., \enquote{Discovery of Algebraic
  Reynolds-Stress Models Using Sparse Symbolic Regression,} {\em Flow,
  Turbulence and Combustion\/}, Vol.~104, 2020, pp.~579–603.

\bibitem{19sw4a}
Sotgiu, C., Weigand, B., Semmler, K., and Wellinger, P., \enquote{Towards a
  general data-driven explicit algebraic Reynolds stress prediction framework,}
  {\em International Journal of Heat and Fluid Flow\/}, Vol.~79, 2019,
  pp.~108454.

\bibitem{16sd2a}
Singh, A.~P. and Duraisamy, K., \enquote{Using field inversion to quantify
  functional errors in turbulence closures,} {\em Physics of Fluids\/},
  Vol.~28, No.~4, 2016, pp.~045110.

\bibitem{16pd2a}
Parish, E.~J. and Duraisamy, K., \enquote{A paradigm for data-driven predictive
  modeling using field inversion and machine learning,} {\em Journal of
  Computational Physics\/}, Vol.~305, 2016, pp.~758--774.

\bibitem{17sm3b}
Singh, A.~P., Medida, S., and Duraisamy, K.,
  \enquote{Machine-Learning-Augmented Predictive Modeling of Turbulent
  Separated Flows over Airfoils,} {\em AIAA Journal\/}, Vol.~55, No.~7, 2017,
  pp.~2215--2227.

\bibitem{11dw2a}
Dow, E. and Wang, Q., {\em Quantification of Structural Uncertainties in the
  $k-\omega$ Turbulence Model\/}, 2011.

\bibitem{16ww3a}
Wu, J.-L., Wang, J.-X., and Xiao, H., \enquote{A Bayesian calibration method
  for reducing model-form uncertainties with application in RANS simulations,}
  {\em Flow, Turbulence and Combustion\/}, Vol.~97, No.~3, 2016, pp.~761--786.

\bibitem{16xw5a}
Xiao, H., Wu, J.-L., Wang, J.-X., Sun, R., and Roy, C.~J., \enquote{Quantifying
  and reducing model-form uncertainties in Reynolds-Averaged Navier–Stokes
  simulations: A data-driven, physics-informed Bayesian approach,} {\em Journal
  of Computational Physics\/}, Vol.~324, 2016, pp.~115--136.

\bibitem{17mp2a}
Meldi, M. and Poux, A., \enquote{A reduced order model based on Kalman
  filtering for sequential data assimilation of turbulent flows,} {\em Journal
  of Computational Physics\/}, Vol.~347, 2017, pp.~207--234.

\bibitem{18mm1a}
Meldi, M., \enquote{Augmented prediction of turbulent flows via sequential
  estimators,} {\em Flow, Turbulence and Combustion\/}, Vol.~101, 2018,
  pp.~389–412.

\bibitem{96tt2a}
Thies, A.~T. and Tam, C.~K.~W., \enquote{Computation of turbulent axisymmetric
  and nonaxisymmetric jet flows using the $k-\epsilon$ model,} {\em AIAA
  Journal\/}, Vol.~34, No.~2, 1996, pp.~309--316.

\bibitem{78ps2a}
Pope, S.~B., \enquote{An explanation of the turbulent round-jet/plane-jet
  anomaly,} {\em AIAA Journal\/}, Vol.~16, No.~3, 1978, pp.~279--281.

\bibitem{91sl2a}
Sarkar, S. and Lakshmanan, B., \enquote{Application of a Reynolds stress
  turbulence model to the compressible shear layer,} {\em AIAA Journal\/},
  Vol.~29, No.~5, 1991, pp.~743--749.

\bibitem{17sm3c}
Shirzadi, M., Mirzaei, P.~A., and Naghashzadegan, M., \enquote{Improvement of
  $k-\epsilon$ turbulence model for CFD simulation of atmospheric boundary
  layer around a high-rise building using stochastic optimization and Monte
  Carlo Sampling technique,} {\em Journal of Wind Engineering and Industrial
  Aerodynamics\/}, Vol.~171, 2017, pp.~366--379.

\bibitem{11co5a}
Cheung, S.~H., Oliver, T.~A., Prudencio, E.~E., Prudhomme, S., and Moser,
  R.~D., \enquote{Bayesian uncertainty analysis with applications to turbulence
  modeling,} {\em Reliability Engineering and System Safety\/}, Vol.~96, No.~9,
  2011, pp.~1137--1149, Quantification of Margins and Uncertainties.

\bibitem{14ec4a}
Edeling, W.~N., Cinnella, P., Dwight, R.~P., and Bijl, H., \enquote{Bayesian
  estimates of parameter variability in the $k–\epsilon$ turbulence model,}
  {\em Journal of Computational Physics\/}, Vol.~258, 2014, pp.~73--94.

\bibitem{16rl4a}
Ray, J., Lefantzi, S., Arunajatesan, S., and Dechant, L., \enquote{Bayesian
  Parameter Estimation of a $k-\epsilon$ Model for Accurate Jet-in-Crossflow
  Simulations,} {\em AIAA Journal\/}, Vol.~54, No.~8, 2016, pp.~2432--2448.

\bibitem{17rl4a}
Ray, J., Lefantzi, S., Arunajatesan, S., and Dechant, L., \enquote{{Learning an
  eddy viscosity model using shrinkage and Bayesian calibration: A
  jet-in-crossflow case study},} {\em ASCE-ASME J Risk and Uncert in Engrg Sys
  Part B Mech Engrg\/}, Vol.~4, No.~1, 09 2017, 011001.

\bibitem{18zf2a}
Zhang, J. and Fu, S., \enquote{An efficient Bayesian uncertainty quantification
  approach with application to $k-\omega-\gamma$ transition modeling,} {\em
  Computers and Fluids\/}, Vol.~161, 2018, pp.~211--224.

\bibitem{16ki3a}
Kato, H., Ishiko, K., and Yoshizawa, A., \enquote{Optimization of Parameter
  Values in the Turbulence Model Aided by Data Assimilation,} {\em AIAA
  Journal\/}, Vol.~54, No.~5, 2016, pp.~1512--1523.

\bibitem{15ky4a}
Kato, H., Yoshizawa, A., Ueno, G., and Obayashi, S., \enquote{A data
  assimilation methodology for reconstructing turbulent flows around aircraft,}
  {\em Journal of Computational Physics\/}, Vol.~283, 2015, pp.~559--581.

\bibitem{17sh6a}
Schaefer, J., Hosder, S., West, T., Rumsey, C., Carlson, J.-R., and Kleb, W.,
  \enquote{Uncertainty Quantification of Turbulence Model Closure Coefficients
  for Transonic Wall-Bounded Flows,} {\em AIAA Journal\/}, Vol.~55, No.~1,
  2017, pp.~195--213.

\bibitem{20rk10a}
Ray, J., Kieweg, S., Dinzl, D., Carnes, B., Weirs, V.~G., Freno, B., Howard,
  M., Smith, T., I.~Nompelis, I., and Candler, G.~V., \enquote{Estimation of
  Inflow Uncertainties in Laminar Hypersonic Double-Cone Experiments,} {\em
  AIAA Journal\/}, Vol.~58, No.~10, 2020, pp.~4461--4474.

\bibitem{16ch2a}
Chowdhary, K. and Najm, H.~N., \enquote{Bayesian estimation of
  Karhunen–Loève expansions; A random subspace approach,} {\em Journal of
  Computational Physics\/}, Vol.~319, 2016, pp.~280--293.

\bibitem{05xi2a}
Xiu, D. and Hesthaven, J.~S., \enquote{High-Order Collocation Methods for
  Differential Equations with Random Inputs,} {\em SIAM Journal on Scientific
  Computing\/}, Vol.~27, No.~3, 2005, pp.~1118--1139.

\bibitem{scikit-learn}
Pedregosa, F., Varoquaux, G., Gramfort, A., Michel, V., Thirion, B., Grisel,
  O., Blondel, M., Prettenhofer, P., Weiss, R., Dubourg, V., Vanderplas, J.,
  Passos, A., Cournapeau, D., Brucher, M., Perrot, M., and Duchesnay, E.,
  \enquote{Scikit-learn: Machine Learning in {P}ython,} {\em Journal of Machine
  Learning Research\/}, Vol.~12, 2011, pp.~2825--2830.

\bibitem{gelmanbda04}
Gelman, A., Carlin, J.~B., Stern, H.~S., and Rubin, D.~B., {\em Bayesian Data
  Analysis\/}, Chapman and Hall/CRC, 2nd ed., 2004.

\bibitem{96gr3a}
Gilks, W.~R., Richardson, S., and Spiegelhalter, D.~J., {\em Markov Chain Monte
  Carlo in Practice\/}, Chapman \& Hall / CRC, Boca Raton, Florida, USA, 1996.

\bibitem{10ll3a}
Liang, F., Liu, C., and Carroll, R.~J., {\em Advanced Markov Chain Monte Carlo
  Methods\/}, Wiley, Chichester, West Sussex, UK, 2010.

\bibitem{Haario2006}
Haario, H., Laine, M., Mira, A., and Saksman, E., \enquote{{DRAM: Efficient
  adaptive MCMC},} {\em Statistics and Computing\/}, Vol.~16, No.~4, 2006,
  pp.~339--354.

\bibitem{07gb3a}
Gneiting, T., Balabdaoui, F., and Raftery, A.~E., \enquote{Probabilistic
  forecasts, calibration and sharpness,} {\em Journal of the Royal Statistical
  Society: Series B (Statistical Methodology)\/}, Vol.~69, 2007, pp.~243--268.

\bibitem{07gr2a}
Gneiting, T. and Raftery, A.~E., \enquote{Strictly proper scoring rules,
  prediction, and estimation,} {\em Journal of the American statistical
  Association\/}, Vol.~102, No. 477, 2007, pp.~359--378.

\bibitem{Zamo2018}
Zamo, M. and Naveau, P., \enquote{Estimation of the Continuous Ranked
  Probability Score with limited information and applications to ensemble
  weather forecasts,} {\em Mathematical Geosciences\/}, Vol.~50, No.~2, 2018,
  pp.~209--234.

\end{thebibliography}
